\documentstyle[aaspp4,epsf,astrobib,11pt]{article}      

\let\oldfootsep=\footnotesep
\def\etal{et~al.}

\def\kmskpc{\,{\rm km \, s^{-1} \, kpc^{-1}}}

\def\msun { \rm {{\em M}_\odot}} 
\def\rsun { \rm {{\em R}_\odot}}
\def\Re{R_{\rm E}}

\def\umin{u_{\rm min}}
\def\ustar{u_{\rm *}} 
 
\def\Rs{R_{\rm *}}
\def\vperp{v_{\rm \perp}}
\def\umin{u_{\rm min}}
\def\t0{t_{\rm 0}}

\def\mlens {M} 
\def\teff{T_{\rm eff}}
\def\vs {{\bf v}_S} 
\def\vl {{\bf v}_L} 
\def\vhbold {\widehat{\bf v}} 
\def\vhat{\widehat{v}} 
\def\that {\widehat{t}}
\def\Msun {\,{\rm M}_\odot} 
\def\kms {\,{\rm km \, s^{-1} }}
\def\kpc {\, {\rm kpc}} 
\def\Mbolsol{M_{\rm bol \odot}}
\def\Mbol{M_{\rm bol}}
\def\mbol{m_{\rm bol}}

\begin{document}

\title{MACHO Alert 95--30 : First Real--Time Observation of Extended
  Source Effects in Gravitational Microlensing} 
\author{
  C.~Alcock\altaffilmark{1,2},           
  W.H.~Allen\altaffilmark{3},            
  R.A.~Allsman\altaffilmark{4},          
  D.~Alves\altaffilmark{1,5,27},         
  T.S.~Axelrod\altaffilmark{6},          
  T.S.~Banks\altaffilmark{7},            
  S.F.~Beaulieu\altaffilmark{8},         
  A.C.~Becker\altaffilmark{2,9},         
  R.H.~Becker\altaffilmark{1,5},         
  D.P.~Bennett\altaffilmark{1,2,10},     
  I.A.~Bond\altaffilmark{11},            
  B.S.~Carter\altaffilmark{3},           
  K.H.~Cook\altaffilmark{1},             
  R.J.~Dodd\altaffilmark{3},             
  K.C.~Freeman\altaffilmark{6},          
  M.~Gregg\altaffilmark{1,5},            
  K.~Griest\altaffilmark{2,12},          
  J.B.~Hearnshaw\altaffilmark{13},       
  A.~Heller\altaffilmark{14},            
  M.~Honda\altaffilmark{15},             
  J.~Jugaku\altaffilmark{16},            
  S.~Kabe\altaffilmark{17},              
  S.~Kaspi\altaffilmark{14},             
  P.M.~Kilmartin\altaffilmark{13,18},    
  A.~Kitamura\altaffilmark{11},          
  O.~Kovo\altaffilmark{14},              
  M.J.~Lehner\altaffilmark{2,12},        
  T.E.~Love\altaffilmark{13,18},         
  D.~Maoz\altaffilmark{14},              
  S.L.~Marshall\altaffilmark{1},         
  Y.~Matsubara\altaffilmark{11},         
  D.~Minniti\altaffilmark{1},            
  M.~Miyamoto\altaffilmark{19},          
  Y.~Muraki\altaffilmark{11},            
  T.~Nakamura\altaffilmark{20},          
  B.A.~Peterson\altaffilmark{6},         
  M.R.~Pratt\altaffilmark{2,9,21},       
  P.J.~Quinn\altaffilmark{22},           
  I.~N.~Reid\altaffilmark{23},           
  M.~Reid\altaffilmark{7},               
  D.~Reiss\altaffilmark{9},              
  A.~Retter\altaffilmark{14},            
  A.W.~Rodgers\altaffilmark{6},          
  W.~L.~W.~Sargent\altaffilmark{23},     
  H.~Sato\altaffilmark{20},              
  M.~Sekiguchi\altaffilmark{19},         
  P.B.~Stetson\altaffilmark{24},         
  C.W.~Stubbs\altaffilmark{2,6,9,21},    
  D.J.~Sullivan\altaffilmark{7},         
  W.~Sutherland\altaffilmark{25},        
  A.~Tomaney\altaffilmark{9},            
  T.~Vandehei\altaffilmark{12},          
  Y.~Watase\altaffilmark{17},            
  D.L.~Welch\altaffilmark{26},           
  T.~Yanagisawa\altaffilmark{11},        
  M.~Yoshizawa\altaffilmark{20},         
  \and
  P.C.M.~Yock\altaffilmark{18}           
\begin{center}
{\bf (The MACHO and GMAN Collaborations) }\\
\end{center}
      }

\altaffiltext{1}{Lawrence Livermore National Laboratory, Livermore, CA 94550
}

\altaffiltext{2}{Center for Particle Astrophysics,
  University of California, Berkeley, CA 94720
}

\altaffiltext{3}{Carter National Observatory, Wellington, New
  Zealand
}

\altaffiltext{4}{Supercomputing Facility, Australian National University,
  Canberra, ACT 0200, Australia 
}
 
\altaffiltext{5}{Department of Physics, University of California, Davis, CA 95616
}

\altaffiltext{6}{Mt.~Stromlo and Siding Spring Observatories,
  Australian National University, Weston, ACT 2611, Australia
}
 
\altaffiltext{7}{Department of Physics, Victoria University,
  Wellington, New Zealand
}

\altaffiltext{8}{Space Telescope Science Institute, San Martin Drive
  3700, Baltimore, MN 21218
}

\altaffiltext{9}{Departments of Astronomy and Physics,
  University of Washington, Seattle, WA 98195
}
 
\altaffiltext{10}{Department of Physics, University of Notre Dame, Notre Dame, IN 46556
}

\altaffiltext{11}{Solar--Terrestrial Environment Laboratory, Nagoya
  University, Nagoya 464-01, Japan
}

\altaffiltext{12}{Department of Physics, University of California,
  San Diego, CA 92093
}
 
\altaffiltext{13}{Department of Physics and Astronomy, University of
  Canterbury, Christchurch, New Zealand
}

\altaffiltext{14}{School of Physics \& Astronomy and Wise Observatory,
   Tel-Aviv University, Tel-Aviv 69978, Israel
}

\altaffiltext{15}{Institute for Cosmic Ray Research, University of
  Tokyo, Tokyo 188, Japan
}

\altaffiltext{16}{Research Institute of Civilization, Tokai University,          
  Hiratsuka 259-12, Japan
}

\altaffiltext{17}{National Laboratory of High Energy Physics (KEK),
  Tsukuba 305, Japan
}

\altaffiltext{18}{Faculty of Science, University of Auckland,
  Auckland, New Zealand
}

\altaffiltext{19}{National Astronomical Observatory, Mitaka, Tokyo
  181, Japan
}

\altaffiltext{20}{Department of Physics, Kyoto University, Kyoto 606,
  Japan
}

\altaffiltext{21}{Department of Physics, University of California,
  Santa Barbara, CA 93106
}
 
\altaffiltext{22}{European Southern Observatory, Karl-Schwarzchild Str. 2, D-85748, Garching, Germany
}

\altaffiltext{23}{Astronomy Department 105-24, Caltech, Pasadena, CA
  91125
}

\altaffiltext{24}{Dominion Astrophysical Observatory,
  5071 West Saanich Road, RR 5, Victoria, BC V8X 4M6, Canada
}
  
\altaffiltext{25}{Department of Physics, University of Oxford,
  Oxford OX1 3RH, U.K.
}
 
\altaffiltext{26}{Departments of Physics and Astronomy,
  McMaster University, Hamilton, Ontario Canada L8S 4M1
}

\altaffiltext{27}{Visiting astronomer, Cerro Tololo Inter-American Observatory
}

\setlength{\footnotesep}{\oldfootsep}
\renewcommand{\topfraction}{1.0}
\renewcommand{\bottomfraction}{1.0}     
\newcommand{\D}{\displaystyle}

\newpage

\vspace{-5mm}
\begin{abstract} 
\rightskip = 0.0in plus 1em

We present analysis of MACHO Alert 95--30, a dramatic gravitational
microlensing event towards the Galactic bulge whose peak magnification
departs significantly from the standard point--source microlensing
model.  Alert 95--30 was observed in real--time by the Global
Microlensing Alert Network (GMAN), which obtained densely sampled
photometric and spectroscopic data throughout the event.  We interpret
the light--curve ``fine structure'' as indicating transit of the lens
across the extended face of the source star.  This signifies
resolution of a star several kpc distant.

We find a lens angular impact parameter $\theta_{min} /
\theta_{source} = 0.715 \pm 0.003$.  This information, along with the
radius and distance of the source, provides an additional constraint
on the lensing system.  Spectroscopic and photometric data indicate
the source is an M4 III star of radius $61 \pm 12 \rsun$, located on
the far side of the bulge at $\sim 9 \kpc$.  We derive a lens angular
velocity, relative to the source, of $21.5 \pm 4.9 \kmskpc$, where the
error is dominated by uncertainty in the source radius.  Likelihood
analysis yields a median lens mass of $0.67^{+2.53}_{-0.46} \msun$,
located with $80 \%$ probability in the Galactic bulge at a distance of
$6.93^{+1.56}_{-2.25} \kpc$.  If the lens is a main--sequence star, we
can include constraints on the lens luminosity.  This modifies our
estimates to $M_{\rm lens} = 0.53^{+0.52}_{-0.35} \msun$ and $D_{\rm
lens} = 6.57^{+0.99}_{-2.25} \kpc$.

Spectra taken during the event show that the absorption line
equivalent widths of H$\alpha$ and the TiO bands near 6700\AA\ vary,
as predicted for microlensing of an extended source.  This is most
likely due to center--to--limb variation in the stellar spectral
lines.  The observed spectral changes further support our microlensing
interpretation.  These data demonstrate the feasibility of using
microlensing limb crossings as a tool to probe stellar atmospheres
directly.
\end{abstract}

\vspace{-5mm}
\keywords{dark matter - gravitational lensing - stars: low-mass, brown dwarfs -
  stars: late-type - stars: atmospheres}

\newpage
\section{Introduction}
\label{sec-intro}

The MACHO collaboration is undertaking an extensive search for
gravitational microlensing by objects in the Galactic halo, bulge, and
disk.  Nightly observations of millions of stars in the Large
Magellanic Cloud (LMC) and Galactic bulge have yielded a total of 8
LMC \cite{macho-lmc2} and more than $120$ bulge events
(\citeNP{macho-bulge1,macho-bulge2}; MACHO Alert
system~\footnote[1]{Current information on the MACHO Collaboration's
  Alert events is maintained at the WWW site {\bf
    http://darkstar.astro.washington.edu}.}).  Similar observations
are being undertaken towards the LMC by EROS \cite{eros} with 2
reported events, as well as towards the bulge by OGLE \cite{ogle} with
18 reported events and DUO \cite{duo} with some 10 reported events.
New efforts to detect the microlensing of unresolved stars in M31 are
being undertaken by the AGAPE \cite{agape} and VATT/Columbia
\cite{vatt} collaborations.  Reviews of the field of gravitational
microlensing are presented in \citeN{gould-rev} and \citeN{pac-rev}.

Statistical analysis of an ensemble of microlensing events provides a
useful discriminant between Galactic structure models.  Two years of
observations towards the LMC yield a microlensing optical depth
representing approximately 50\% of a ``standard'' Galactic halo
comprised entirely of massive compact halo objects \cite{macho-lmc2}.
Similar analysis of Galactic bulge data, as reported by MACHO and OGLE
\cite{macho-bulge2,ogle-tau}, indicate an optical depth a factor of
$\sim 3$ larger than predicted by axisymmetric Galactic models (e.g.
\citeNP{macho-griest91,pac91}).  Consistent Galactic models have
recently been constructed which include a Galactic bar viewed nearly
end--on (\citeNP{kir-pac,bar,binney}).  However, it is difficult to
identify clearly a lensing population because the mass, velocity, and
distance of each lens are not uniquely determined in the standard
microlensing solution.

Gravitational microlensing is characterized by the transient,
achromatic brightening of a background star due to gravitational
deflection of its light by a massive ``lens'' passing near our line of
sight to the source.  This results in the distortion of the source
disk into multiple (unresolved) images whose total brightness is
greater than that of the original source brightness.  Given the
assumptions of a point--mass lens, point source, and unaccelerated
motion, the source magnification $A(t)$ is of the simple form

\begin{eqnarray}
\label{eq-amp}
 & A(t) = \frac{\D u^2 + 2}{\D u\sqrt{u^2 + 4}}, & \\
 {\rm where} & u(t) = \sqrt{u_{min}^2 + (~2(t-t_0)/\that~)^2} . \nonumber &
\end{eqnarray}

These equations describe the event's ``light--curve'' as a function of
time.  $A(t)$ is the observed source magnification, $\that = 2
\Re/\vperp$ is the characteristic event timescale, where $\vperp$ is
the transverse velocity of the lens, and

\begin{eqnarray}
\label{eq-re}
 & \Re^2 \equiv \frac{\D 4G \mlens }{\D c^2}x(1-x) L &
\end{eqnarray}
is a conveniently defined distance scale in the lens plane (the lens's
Einstein radius).  $\mlens$ is the mass of the deflector, and $x$ is
the ratio of the observer--lens to observer--source ($L$) distances.
The equation for $u(t)$ describes the impact parameter of the lens,
scaled by $\Re$, as it passes near the source, and $\t0$ and $\umin$
are the values of $t$ and $u$ at peak magnification.
        
Fitting a normal microlensing event provides only one parameter,
$\that$, containing information about the lens.  However, $\that$ is a
function of the three unknowns $\mlens$, $x$ and $\vperp$. (Strictly
speaking, the source distance $L$ is also unknown, but the uncertainty
in $L$ is typically much smaller than that in $x$, so it is simpler to
treat $L$ as known).  This indicates a continuum of event parameters
which can conspire to produce similar duration microlensing events.


\section{Light--Curve Fine Structure and Extended Source Effects}
\label{sec-finestructure}

In instances where the standard approximations break down,
light--curve ``fine structure'' describes deviations from the standard
model whose nature may break this degeneracy.  Important microlensing
events in this regime include the detection of binary microlensing
\cite{ogle7,macho-pratt,duo2,macho-causitc}, and the observation of
parallax in a gravitational microlensing event \cite{macho-parallax}.
An exciting possibility is the detection of short timescale deviations
in an event light--curve due to the presence of planets in the ``lensing
zone'' of the microlensing system \cite{mao-pac,gould-loeb,ben-rhie}.

The effects of extended source size may also become apparent in the
limit of a large source, high magnification event, where the lens
impact parameter is of the order of the projected source radius
\cite{gould-pm,nemiroff94,witt-mao,peng}.  In this situation, an extra
parameter is included in the microlensing solution, $\ustar = x \Rs
/\Re$.  Here $\Rs$ is the radius of the source star, and $\ustar$
describes its projection into the lens plane, scaled by $\Re$.  Due to
its geometric nature, $u(t)$ is not changed by this parameterization.
An analytic solution for $A(t)$ in this limit is presented in
\citeN{witt-mao}.

Fitting the light--curve with the extra parameter $\ustar$ provides a
second constraint equation for the three lens parameters - the angular
size of the source in terms of the lens's Einstein radius.  If we
estimate the linear radius and distance of the source star from
photometry and spectroscopy, we have an estimate of the lens proper
motion relative to the source, which leads to a unique mass--distance
relation for the lens.  Such a situation is rare, and would be
expected in only $\sim 5\%$ of events towards the galactic bulge
\cite{gould-welch}.

To account for limb--darkening of the source, an appropriate
limb--darkening law and coefficients must be determined for the source
star.  The magnification is then integrated over the face of the star,
properly weighted by the brightness profile.  The limb--darkening law
we will choose is of the form (e.g. \citeN{claret-r})

\begin{eqnarray}
\label{eq-limbd}
I_\lambda(\mu)/I_\lambda(1) = 1 - a(1 - \mu) - b(1 - \mu)^2. 
\end{eqnarray}

Here $\mu$ is the cosine of the angle between the observer's line of
sight and the emerging stellar radiation, and $a$ and $b$ are model
parameters dependent primarily upon the effective temperature and
surface gravity of the star.

\section{Global Microlensing Alert Network (GMAN)}
\label{sec-gman}

To provide both the temporal and photometric resolution necessary to
distinguish light--curve ``fine structure'', coordinated follow--up
observations are being undertaken through the Global Microlensing
Alert Network (GMAN) \cite{macho-pratt,macho-alert1}, as well as the
PLANET collaboration \cite{planet}.  

GMAN utilizes telescope resources at Cerro Tololo Inter-American
Observatory~\footnote[2]{Operated by the Association of Universities
  for Research in Astronomy, NOAO, under cooperative agreement with
  the National Science Foundation} (CTIO), Mount John University
Observatory (MJUO), University of Toronto Southern Observatory (UTSO),
and Wise Observatory which respond nightly to the dynamic status of
ongoing microlensing events (see Table~\ref{GMAN}).  Observations from
MJUO are carried out by the MOA collaboration of Japan and New
Zealand, which is undertaking a general survey of microlensing events
towards the Magellanic Clouds and the Galactic bulge \cite{moa}.  The
observational flexibility available at these locations provided nearly
continuous coverage of this event.

Microlensing follow--up images are automatically processed on site.
Photometry is carried out using IRAF~\footnote[3]{IRAF is distributed
by the National Optical Astronomy Observatories, which is operated by
the Association of Universities for Research in Astronomy, Inc., under
contract to the National Science Foundation.} scripts which call
DaophotII \cite{daophot}.  Reductions are usually finished within 6
hours of image acquisition.  Team members have immediate network
access to GMAN data, which were subsequently used for real--time
scheduling during this event.  Normalization is performed on--the--fly
using a list of reference stars obtained from the MACHO data set.  The
reported DaophotII error bars for UTSO and WISE photometry have been
increased by a factor of 2.0, to account for the typical scatter in a
time series of constant stars from these images.  The MJUO data used
in this analysis were obtained manually from the raw frames after the
event, using DaophotII operating within the IRAF environment; the
uncertainties were estimated by examining the variations of standard
star ratios within and between frames.  Data from CTIO were
re--reduced using Allframe \cite{allframe}.

The system and data collection of the MACHO experiment are described
in \citeN{macho-lmc2}.  The data are obtained simultaneously in two
specially designed wide--pass filters with effective wavelengths of
approximately 5150 \AA\ and 6900 \AA.  Transformations from the MACHO
instrumental photometry system to the standard $V$ and $R_{KC}$ system
have been derived by comparisons with several thousand tertiary
standard stars calibrated in the LMC.  Checks of our photometry with
published photometry in Baade's Window confirm a zero point
uncertainty of less than 0.1 mag in $V$ and $R$, where this
uncertainty is a conservative estimate of small zero point differences
between fields monitored by MACHO in the bulge.

\section{Photometry of MACHO Alert 95--30}
\label{sec-event}

The source star in Alert 95--30 is located at $\alpha = $18:07:04.26,
$\delta = -$27:22:06 (J2000).  The MACHO project's identifier for this
star is 101.21821.128.  Alert 95--30 was detected by the MACHO Alert
system at $A \sim 1.8$ on Jul 24, 1995, approximately 22 days before
the observed peak.    

The location of the source star in an optical color--magnitude diagram
(CMD), $V$ = 16.21 and $(V-R)$ = 1.39, indicates this star passes the
``clump giant'' cuts defined in \citeN{macho-bulge2}.
Figure~\ref{fig-cmd} shows a single epoch CMD for a 2$\times$2 arcmin
field surrounding the source star, which is indicated with a circle.
We may be reasonably sure then that this star is located in the
Galactic bulge and has a large radius.  Microlensing fits to the
rising portion of the light--curve predicted a high--magnification
event for the giant source, which presented the possibility of source
star resolution.

Subsequent observations indicated deviations from the standard
microlensing model at $A \sim 20$, approximately 2 days before the
projected peak.  At this time, we mounted an aggressive program of
photometry and spectroscopy to study these deviations thoroughly.
Follow--up observations continued nightly past the observed peak until
the star returned to its baseline state.  Intermittent observations
followed to determine accurately the baseline flux as measured at each
site.

Figure~\ref{fig-all} displays the MACHO 1995 bulge season light--curve
of Alert 95--30~\footnote[4]{We report an uncatalogued CCD trap which
allowed several contaminated MACHO--red observations to pass through
our processing stream.  The majority of these were removed from the
data set based upon their proximity to the bad columns and psf FWHM.}.
The alert date is indicated with an arrow, after which time the
light--curve becomes heavily sampled with follow--up data.  Data from
all follow--up observatories have been included after determining the
baseline flux in each passband.  No microlensing fits are included in
the figure, but it is apparent that the data conform to the symmetric
and achromatic shape expected of gravitational microlensing events, at
least for magnification $A$ $\le$ 20.

This conclusion is confirmed after fitting a normal microlensing curve
to the combined dataset.  Event parameters for this fit (fit 1
hereinafter) are listed in Table~\ref{tab-stat1}, and $\chi ^2$
statistics for each passband are listed in Table~\ref{tab-stat2}.
However, near the peak of the event the data deviate from the expected
light--curve (see Fig~\ref{fig-peak}).  This type of deviation is not
unexpected in a high magnification microlensing event.  We note that
the standard point--source microlensing model allows infinite
magnification when the source and lens are aligned.  When the extended
size of the source is considered, the magnification is limited, as the
entirety of the source disk cannot be perfectly aligned with the lens.
If we include an extra parameter in the fit to account for the angular
size of a constant surface brightness source star (fit 2), we reduce
$\chi ^2$ by 1084.  The deviations near the peak of MACHO 95--30 are
significantly reduced with this model, which we interpret as clear
justification for the extended source microlensing interpretation.

The data in Table~\ref{tab-stat2} indicate an extended source fit
$\chi^2$ per degree of freedom of approximately 1 for the GMAN
follow--up data.  The implication here is that the data scatter around
the fit is what one would statistically expect due to measurement
error.  Thus, the data are in excellent agreement with the extended
source microlensing fit, considering our photometry is at the $1-2\%$
level, as indicated in column 3 of Table~\ref{tab-stat2}.  However,
the MACHO data exhibit $\chi^2_{dof}$ between $3 - 4$, which formally
indicates a poor fit to the data.  While it is obvious that this is
indeed a gravitational microlensing event, such a poor fit might cast
doubt on the extended source interpretation.  Fits to data of
magnification $A < 10$, which should be negligibly affected by
extended source effects, display the same excessive scatter around the
fit.  We therefore conclude that the majority of this scatter is
contained in baseline measurements, and should not adversely affect
our interpretation of the data.

Figure~\ref{fig-all} also presents a scaled schematic of this
microlensing event using event parameters derived from fits with a
limb--darkened source star (fit 3).  Included are the lens's Einstein
radius expressed as $\that / 2$, the source radius scaled to 0.0756
$\Re$, and the trajectory of the lens across the extended face of the
source star.

Figure~\ref{fig-peak} provides a visual comparison between the best
standard microlensing fit to the data (dashed line) and microlensing
of a limb--darkened extended source (solid line).  Each passband was
fit independently with a baseline parameter, while event parameters
were derived from joint fits on the combined dataset.  All fits were
performed in the MINUIT environment \cite{minuit}.

\section{Determination of Stellar Parameters}
\label{sec-stellar}

Analysis of the stellar parameters of the source star in MACHO 95--30
(hereinafter referred to as the source star) is required if we are to use
the additional information provided by $\ustar$ to characterize the
lensing object.  The most important parameters are its distance ($L$)
and radius ($\Rs$).  Also valuable are estimates of the source
temperature and surface gravity, which will help define the
limb--darkening coefficients.

Figure~\ref{fig-cmd} shows the fiducial line for the red giant branch
of the metal--rich globular cluster NGC~6553 from \citeN{ortolani},
plotted over the optical CMD for a 2$\times$2 arcmin field surrounding
the source star.  We note that the horizontal branch (HB) in this
field is inclined like that of NGC~6553, due to the effect of
differential reddening.  Even though the morphology of the red giant
branch of this cluster is similar to that of the MACHO field
surrounding the microlensing event, the magnitude of the source star
is fainter than expected, after we account for the different
reddenings and distances to these sources (we adopt a bulge distance
of 8 kpc).  The smaller apparent luminosity could be due to
differences in age, metallicity, or distance.  From this diagram, we
tentatively conclude that this star is either more metal-rich, or more
distant than 8 kpc (or both) than the giants of similar color in
NGC~6553.

\subsection{Reddening}
\label{sec-reddening}

The reddening is patchy in the MACHO field where the source star is
located.  Figure~\ref{fig-rrl} shows the amplitude vs. color of all
the RR Lyr that we found in this field.  The reddening vector is
horizontal in this figure, so the color spread is mostly due to
differential reddening within the field, as the RRab sequence in such
a diagram should be very tight \cite{bono}.  There appear to be two
zones of different obscuration, with $E(B-V)$ ranging from 0.3 to 0.8
mag.  The source star appears to be located in the region with smaller
reddening, as measured from the colors of the three closest RR Lyr.
We identified a total of two bulge RR Lyr type ab, and one RR Lyr type
c, within 2 arcmin of the source star.  The locations of these 3 RR
Lyr are shown in the optical CMD (Figure~\ref{fig-cmd}) as triangles.
The mean color of the two RRab is $V-R = 0.54$, and the mean color of
the RRc is $V-R = 0.46$.  From these colors we deduce $E(V-R)=0.34 \pm
0.05$. Using the extinction law of \citeN{rieke}, we obtain
$A_V=1.36$, $E(B-V)=0.44$, $A_K=0.15$, $E(V-K)=1.21$, $E(J-H) = 0.15$,
$E(H-K) = 0.08$, and $E(J-K)=0.23$.  The observed and dereddened
photometric measurements of the source star are listed in
Table~\ref{tab-phot1}.

\subsection{Infrared Photometry}
\label{sec-irphot}

We observed the field of the source star with the CIRIM camera on the
CTIO 1.5m telescope.  The f/13.5 configuration yields a pixel scale of
0.65 arcsec pixel$^{-1}$.  The detector is a 256$\times$256 HgCdTe
NICMOS 3 array.  Conditions on the night of September 24, 1996 were
photometric.  We used the $J$, $H$, and $K_{s}$ filters, and a
five-position on-source dithering pattern optimally to expose the
target field and collect the necessary frames for sky flats.  Six
10--second coadds were obtained at each position in the dithering
pattern for total exposures of 300 seconds in each filter.  The
individual IR images were reduced following standard procedure using
dark and sky frames, then shifted and combined to create the final
images, which cover about 2 arcmin on a side.  The photometric
calibration was obtained by observing the UKIRT Faint Standard Star
no.~1 (G158--100) just prior to observing the field of the source
star.  The standard star and program field were observed at identical
airmasses ($X=1.2$), thus no extinction corrections were necessary.
We find $K$ = 9.98, $(J-K)$ = 1.12, and $(H-K)$ = 0.26 mag for the
source star.  We estimate an uncertainty of 0.05 mag in $J$, $H$, and
$K$ from the photon statistics and aperture corrections.
Figure~\ref{fig-ccd} shows the optical--IR color--color diagram of
stars one magnitude brighter than the horizontal branch in the
2$\times$2 arcmin field surrounding the source star, along with the
sequences corresponding to field giants, field dwarfs, and Baade's
Window giants from \citeN{frogel-whit}.  The colors of the source star
are similar to Baade's Window M giants, and we conclude that it is
unlikely to be a field dwarf.

Figure~\ref{fig-oircmd} shows an expanded version of the reddening
corrected $K_{0}$ vs $(V-K)_{0}$ optical--IR CMD for the bright stars
found in this field.  Overplotted are the fiducial lines corresponding
to the red giant branches of the metal--rich globular clusters 47~Tuc
from \citeN{frogel} and the metal-rich globular cluster NGC~6553 from
\citeN{guarnieri}, assuming a distance of 8 kpc for both populations.
Figure~\ref{fig-oircmd} shows again that the source star is relatively
faint with respect to the giants in these clusters. This suggests that
the star is either more metal-rich than $[Fe/H] = -0.4$, and/or more
distant than 8 kpc.  As an additional comparison, we have plotted data
from \citeN{frogel-whit} for spectral type M3--5 giants from Baade's
Window (assuming a distance of 8 kpc), and our own estimated fiducial
line for these stars.  If we assume a solar composition for the source
star ($[Fe/H] = 0$), we obtain a distance of about $D = 10$ kpc.  The
uncertainty in this distance is large; it assumes that all of the
scatter about the fiducial line of BW giants is due to the
line--of--sight depth of the bulge.  It is possible that the source
star is on the metal--rich tail of the distribution of bulge giant
metallicities.  Comparing the two globular cluster tracks in
Figure~\ref{fig-oircmd} gives one a sense of the dependence of the red
giant branch fiducial lines on metallicity.  If the location of the
source star in the CMD were entirely due to metallicity, it would be
an extremely metal--rich star, which is less probable than a
combination of distance and metallicity effects at work.  Given the
position of the source star in the CMD at the outer envelope of giants
observed in Baade's Window, we conclude that the source star is likely
located on the ``far'' side of the bulge.  Note that we would expect a
giant star in the Sagittarius dwarf galaxy to be about 2 magnitudes
fainter than the observed $K_{0}$ magnitude of the source star (see also
Section~\ref{sec-radvel} below).

\subsection{Spectroscopy}
\label{sec-spec}

We have obtained spectra of the source star in several different
observing runs, as will be discussed in Section~\ref{sec-spectra}.
Here we analyze the spectra taken at the 4m Blanco telescope at CTIO
with the R--C spectrograph and the Loral~$3K_1$ detector during the
nights of 14 and 15 August, and 27 September, 1995.  The first two
spectra were taken 0.5 days before and after, respectively, the peak
of the microlensing event, while the lens was still in transit across
the source face.  The wavelength coverage of these spectra is from
$\lambda 6230$ to $\lambda 9340$ \AA.  A HeNeAr comparison lamp was
used, which provided the wavelength calibration.  The exposures were
60 seconds long, and were taken at an airmass of 1.7.  All the spectra
were reduced in the standard way.  The dispersion was 1\AA\
pixel$^{-1}$, and the resolution was 5\AA, as measured from the FWHM
of the lines.

The additional spectrum from 27 September, 1995 was taken at an
airmass of 1.1, with a total exposure time of 500 seconds.  No independent
wavelength calibration was obtained for this spectrum.  The dispersion
was 2\AA\ pixel$^{-1}$, and the resolution was 8\AA.  The wavelength
coverage of this spectrum, shown in Figure~\ref{fig-spec1}, was from
$\lambda 3890$ to $\lambda 9830$ \AA.  This spectrum, taken 42.3 days
after peak magnification, is very similar to the near--peak spectra.
This further supports the microlensing interpretation, where the
overall colors or spectral type must not change.

\subsubsection{Spectral Typing}

We determined the spectral type of the source star by direct
comparison with the digital spectral atlas of late-type stars of
\citeN{turnshek} and \citeN{kirkpatrick}.  This comparison was done
visually after rebinning our CTIO spectra to lower resolution, and
paying close attention to the match of the prominent TiO bands.  There
is a very close match of the spectra shown in Figure~\ref{fig-spec1}
with a spectral type M4 III, and we adopt this classification for the
source star.  The uncertainty in this classification is estimated to
be one spectral subtype.

Using the spectral type we can check our estimate of the reddening of
the source star by adopting the unreddened optical and infrared colors
of the same type bulge giants.  For an M4 III bulge star,
\citeN{frogel-whit} give $(J-K)_0 = 0.90$ and $(H-K)_0 = 0.18.$ and
from the observed infrared colors we compute $E(J-K) = 0.22$, and
$E(H-K) = 0.08$, respectively.  These reddening values are in
excellent agreement with the reddening derived from the optical
photometry of the RR Lyr in this field (see Table~\ref{tab-phot1}).

\subsubsection{CaII Triplet}

The CTIO spectra allows us to determine independently the surface
gravity ($g$) of the source star.  The spectral region covered
includes the CaII triplet, which is strong and very sensitive to log
$g$. Other spectral lines in the region from $\lambda 6400$ \AA\ to
$\lambda 9200$ \AA\ can be used for this purpose in M--type stars,
like the NaI doublet at $\lambda\lambda 8183,8195$\AA\ 
\cite{kirkpatrick}.  However, the lines of the CaII triplet are the
most suitable ones, because they have been studied and calibrated by a
number of authors
(\citeNP{jones,humphreys,diaz,alloin,jorgensen,erdelyi}).

We measured the equivalent width of the CaII lines at $\lambda
8498.06$ \AA, $\lambda 8542.14$ \AA, and $\lambda 8662.17$ \AA\ in the
three spectra obtained with two different resolutions, obtaining $EW =
11.0 \pm 0.5$ \AA, where the uncertainty is dominated by the location
of the continuum.  We used the local continuum bands at $\lambda
8480$ \AA, $\lambda 8635$ \AA, and $\lambda 8905$ \AA, defined by
\citeN{jones}, \citeN{diaz}, and \citeN{alloin}, in order to make a
direct comparison with their measurements and models.  Note that the
resolution of our spectra is similar to these other works.

We have also compiled published data (spectral type, $\teff$, log $g$, and
CaII$-EW$) on M--type stars, covering a wide range of temperatures and
gravities.  In some cases only the two strongest lines of the CaII
triplet are measured, and we have scaled these accordingly in order to
include the weaker line.  Figure~\ref{fig-ew1} shows the equivalent
width vs M spectral subtype for dwarfs (V), giants (III), and
supergiants (I).  The position of the source star is shown with the
box.  This figure confirms the conclusions drawn from the IR
photometry and spectral classification $-$ the source is a giant star.

Figure~\ref{fig-ew2} shows equivalent width vs log~$g$.  The dotted
lines show the $\pm 1 \sigma$ measurements for the source star.  From
this figure we conclude that log~$g$ $= 1.0 \pm 0.2$.  A similar value
is obtained by applying the empirical relations found by
\citeN{alloin} and \citeN{diaz}, and by slightly extrapolating the
models of \citeN{jorgensen} and \citeN{erdelyi}.

\subsubsection{Radial Velocity}
\label{sec-radvel}

We have also measured the radial velocity of the source star.
Unfortunately, we did not measure radial velocity standards, and have
to rely on a wavelength calibration based on the HeNeAr lamps.  From
the two CTIO spectra of the August run we obtain $V_r = -80 \pm 5$ km
s$^{-1}$.  Even though the spectra from the August run have high
enough resolution to measure velocities good to $< 10$ km s$^{-1}$, we
consider that the most important source of uncertainty is the zero
point.

We consider the possibility that the source star may be located in the
Sgr dwarf galaxy.  The Sgr dwarf galaxy discovered by \citeN{ibata} is
much more extended than previously thought.  RR Lyr belonging to this
galaxy have been identified in fields close to the Galactic plane
(\citeNP{alard-sgr,macho-sgr}).  In fact, the Sgr galaxy extends as
far as the field containing the source star, as shown by the one Sgr
RR Lyrae type ab discovered in this field by \citeN{macho-sgr}.
However, the radial velocity measured here rules out membership to the
Sgr dwarf galaxy, which has $V_r = +150$ km s$^{-1}$.

The radial velocities of bulge giants in off-axis bulge fields have
been measured by \citeN{minniti} and \citeN{minniti-liebert}.  At the
position of the MACHO field, the mean heliocentric radial velocity of
bulge stars is estimated to be about 25 km s$^{-1}$, with a velocity
dispersion $\sigma = 80$ km s$^{-1}$.  The radial velocity of the
source star is, therefore, consistent with the velocities of bulge
giants, supporting our assumption that this star belongs to the bulge
population.

\subsection{Stellar Parameters}
\label{sec-parameters}

We will derive the stellar parameters $M$, $\teff$, $L$, and $R$ for
the source star in a variety of ways and compare the different values to
estimate our uncertainties.  When needed, we assume the solar values:
$T_{\odot} = 5730 K$, log~$g_{\odot} = 4.44$, and $\Mbolsol =
4.72$ mag.  The following equations relate various stellar parameters:
$$(L/L_{\odot}) = (R/R_{\odot})^2 (\teff/T_{\odot})^4$$
$$(g/g_{\odot}) = (M/M_{\odot}) (R/R_{\odot})^{-2}$$

We begin with the most simple method, adopting the typical stellar
parameters of an M4 giant in the galactic bulge.  We would expect the
mass of the source star to be $\sim 1 M_{\odot}$, the mass appropriate
for a bulge giant according to the estimated age of the bulge (e.g.
\citeNP{holtzman}).  Using only the spectral type information (M4
III), we obtain $\teff = 3430$, $\Mbol = -2.7$ from \citeN{lang},
which gives $R = 85 R_{\odot}$.  Also, for an M4 III star, $R=83
R_{\odot}$, and $\teff= 3600$ from the recent calibrations of
\citeN{dyck}.  These values are averages over several spectral
sub-types.  The dependence of radius on spectral type is such that for
an M3 III star $R \approx 60 R_{\odot}$, and for an M5 III star $R
\approx 100 R_{\odot}$.

The IR colors allow us to measure the effective temperature.  For
$J-K_{0} = 0.89$, $\teff = 3900 \pm 270$ from the calibration
of \citeN{feast}.  For $V-K_{0} = 5.03$, $\teff = 3600$ from the
recent calibration of \citeN{bessell2}.  The optical photometry also
allows us to estimate the temperature of the source star.  For
$(V-R)_0 = 0.93$, we find $\teff = 3800 \pm 200$ by differential
comparison with the giant star III-17 member of the globular cluster
NGC~6553 analyzed by \citeN{barbuy}.  Comparing these values of
temperature with those derived from the spectral type information, we
adopt a temperature of $\teff = 3700 \pm 250$ for the source star, or
log$(T/T_{\odot}) = -0.19 \pm 0.03$.

Numerous bolometric corrections exist in the literature.  We find
$BC_K(J-K) = 2.6$ and $BC_K(V-K) = 2.7$ from \citeN{frogel-whit} and
\citeN{bessell2}, $BC_H(M~{\rm giants}) = 2.6$ from \citeN{bessell},
and $BC_V(M4 III) = -2.2$ \cite{lang}.  These give $\mbol =$ 12.4,
12.5, 12.6, and 12.6, respectively.  We adopt $\mbol = 12.5 \pm 0.1$
for the source star.

In order to compute the absolute magnitudes, we must assume a distance
to the star.  Adopting a distance of 8 kpc, or $m-M_0 = 14.52$, we
obtain the absolute magnitudes $M_V=+0.34$, $M_K=-4.69$, and $\Mbol =
-2.1$ mag.  This gives log$(L/L_{\odot}) = 2.69 \pm 0.04$.  We can
immediately derive the radius, log$(R/R_{\odot}) = 1.72 \pm 0.06$ or
$R = 53 \pm 8 R_{\odot}$.  However, the uncertainty here is
underestimated given the systematic uncertainty in the adopted
distance.  From our arguments in Section~\ref{sec-irphot} , we adopt
the distance to the source star of $9 \pm 1$ kpc, giving a distance
modulus $m-M_0 = 14.77 \pm 0.25$.  Re--calculating the radius at this
distance and incorporating the added uncertainty, we find $\Mbol =
-2.25 \pm 0.27$ and thus $R = 61 \pm 12 R_{\odot}$.

Lastly, we derive the mass of the source star using the measured value
of log $g$ (Section~\ref{sec-spec}).  We find log$(g/g_{\odot}) =
-3.44 \pm 0.08$ and $M \sim 0.8 - 2.5 M_{\odot}$.  The mass derived in
this manner is quite uncertain.  The largest mass values are unlikely
given the age of the bulge $t \approx 10$ Gyr \cite{holtzman}.
However, a mass slightly larger than 1 $M_{\odot}$ is consistent with
the source star having a high metallicity.

To summarize, we adopt the following stellar parameters for the source
star in MACHO Alert 95--30, which we categorize as an M4 bulge giant:

$$L = 600 \pm 200 L_{\odot}, $$
$$\teff = 3700 \pm 250 K, $$
$$log ~g = 1.0 \pm 0.2, $$
$$R = 61 \pm 12 R_{\odot}, $$
$$D = 9 \pm 1~kpc, $$
$$M \approx 1.0 M_{\odot}, $$
$$[Fe/H] \approx 0. $$ 

\section{Determination of Event Parameters}
\label{sec-params}

\subsection{Effect of Limb--Darkening on the Photometry}
\label{sec-limbd}

We now integrate an approximate limb--darkening law into the
microlensing model, to extract more realistic event parameters than
those derived with a constant surface brightness disk.
Limb--darkening coefficients for GMAN's standard R and V passbands
exist in the literature for $\teff$ = 3500 K, log $g$=1 for the
quadratic form of Equation~\ref{eq-limbd} \cite{claret-r,claret-v}.
Limb--darkening coefficients were also calculated for MACHO's
non--standard passbands \cite{claret-macho}.  These coefficients are
listed in Table~\ref{tab-limbd}.

Including this brightness profile in the extended source model further
improves the fit $\chi ^2$ by 9 (see Table~\ref{tab-stat2}, fit 3).
While this improvement is formally significant, its interpretation
here appears unclear.  Comparison of fits 2 and 3 in
Table~\ref{tab-stat2} shows no clear trend between data sets, which
effectively washes out any overall conclusions about the significance
of the model.  However, we do regard the limb--darkened parameters as
the more realistic interpretation of the data.  Comparison of optical
and {\it infrared} photometry during a lens transit should detect
significant color terms as a result of source limb--darkening
\cite{gould-welch}, which would then provide a more robust test of
stellar atmosphere models.

\subsection{Implications for Lens Mass} 
\label{sec-mass} 

The limb--darkened fit to finite source microlensing provides a direct
measurement of $\ustar \equiv \theta_* / \theta_E = 0.0756$, i.e. the
angular size of the star as a fraction of the Einstein ring angle.  To
convert this into physical units, we need an estimate of the angular
radius $\theta_*$ of the source star.  The analysis in
Section~\ref{sec-stellar} indicates $R_* = 61 \rsun$ and $L = 9 \kpc$,
giving $\theta_* = 31.5 \,\mu{\rm arcsec}$ and thus $\theta_E = 0.417
\, {\rm milliarcsec}$.  The Einstein diameter crossing time $\that$ is
measured similarly to the usual point--source case, so this provides
the proper motion of the lens with respect to the source star $\omega
\equiv 2 \theta_E / \that = 21.5 \kmskpc = 4.51$ mas/yr.

Thus, an extended source microlensing event provides us with 2
constraints $\that$ and $\omega$ on the 3 unknowns $\mlens$, $x$ and
$v_\perp$ of the lens.  In the following, it is convenient to define
$\vhat \equiv \omega L = v_\perp / x$ to be the relative velocity of
the lens projected back to the source plane (cf \citeNP{han-spec});
substituting $r_E = \vhat x \that / 2$ into Eq.~\ref{eq-re}, we obtain
a one-to-one relationship between the lens mass and distance,
\begin{equation} 
\label{eq-mx} 
 M(x) = {\vhat^2 \that^2 c^2 \over 16 \, G L } { x \over 1-x } . 
\end{equation} 
(By symmetry, this is the same as Eq.~6 of \citeN{macho-parallax}
exchanging observer and source, i.e. with $\vhat$ instead of
$\tilde{v}$, and $x \leftrightarrow 1-x$.)  For the observed values of
$\vhat = 193 \kms$, $\that = 67.3$ days, we have $M(x) = 0.192 \Msun
\, x/(1-x)$, thus the lens may be either a low-mass star roughly
half--way to the source, or a more massive star closer to the source.
Figure~\ref{fig-like} shows $M(x)$ for the event parameters given in
Table~\ref{tab-stat2} (fit 3), and the radius and distance of the
source star obtained in Section~\ref{sec-stellar}.  Since our proper
motion error is dominated by uncertainty in the source radius, we also
include $M(x)$ contours for $\Rs = 61 \pm 12 R_{\odot}$.

We may obtain additional constraints on $x$ by using a model for the
distributions of sources and lenses, since the likelihood of obtaining
the observed value of $\vhat$ is also a function of $x$.

For given lens mass, the rate of microlensing is proportional to 
\begin{equation}
\label{eq-gamma}
d\Gamma \propto \sqrt{x (1-x)} \, \rho_L (x) \, v_\perp f_S(\vs) \,
f_L(\vl) \, dx \, d\vs \, d\vl .
\end{equation}
where $\rho_L$ is the density of lenses at distance $x$, $f_L(\vl)$
and $f_S(\vs)$ are the lens and source velocity distribution functions
(normalized to unity) in the plane perpendicular to the line of sight.
The source and lens velocities $\vs, \vl$ are related to $\vhbold$ by
$\vl = (1-x) {\bf v}_\odot + x (\vs + \vhbold)$, where $\vhbold =
(\vhat \cos \phi, \vhat \sin \phi)$ and $\phi$ is the (unknown)
direction of the relative proper motion.

Given a model for $\rho_L, f_S, f_L$, we may integrate
Eq.~\ref{eq-gamma} and thus obtain joint probability distributions for
any of the variables.  Since we have measured $\vhat$, and the lens
mass depends on only one unknown ($x$), we need to consider the joint
probability distribution of events in the $(x, \vhat)$ plane, and then
marginalize to get a probability distribution of $x$ given the observed
value of $\vhat$.

Thus, we change variables from $d\vl$ to $d\vhat \,
d\phi$, giving 
\begin{equation}
\label{eq-gamma2}
d\Gamma \propto \sqrt{x (1-x)} \, \rho_L (x) \, v_\perp f_S(\vs) \,
f_L(\vl) \, dx \, d\vs \, 
\left| {\partial \vl \over \partial(\vhat,\phi)} \right|
d\vhat \, d\phi.
\end{equation}

We then substitute for $\vl$ and $v_\perp$ and integrate over the
unknowns $\vs$ and $\phi$, giving a likelihood as a function of
distance for given $\vhat$,
\begin{equation} 
\label{eq-like}
 {\cal L}(x;\vhat) = \left.{d\Gamma \over dx d\vhat}\right|_{\vhat}  \propto 
 \sqrt{x (1-x)} \, \rho_L (x) \, \vhat^2  \, x^3 
 \int \, \int f_S(\vs) 
 \, f_L\left( (1-x) {\bf v}_\odot + x (\vs + \vhbold ) \right) 
 \, d\vs \, d\phi,  
\end{equation}

This result may be understood as follows: the integrals are over all
combinations of source and lens velocity which give rise to the
observed $\vhat$.  The $\sqrt{x(1-x)}$ and $\rho_L$ terms arise from
the $x-$dependence of the Einstein radius and density of lenses,
respectively.  There is a factor of $v_\perp = \vhat x$ because a
given lens contributes a lensing rate $\propto v_{\perp}$, and a
factor of $x^2 \vhat $ from the Jacobian $\det( \partial\vl
/\partial(\vhat,\phi) )$.

To evaluate Eq.~\ref{eq-like}, we adopt a disk velocity dispersion of
$30 \kms$ in each direction, with a flat rotation curve of $220 \kms$.
We adopt a bulge velocity dispersion of $80 \kms$ in each direction,
and no bulge rotation.  For the density profiles, we use a standard
double--exponential disk, and a barred bulge as in \citeN{han-bar}.
The source is located at galactic coordinates $l = 3.73^o, b =
-3.30^o$, and we assume $L = 9 \kpc$, $R_0 = 8 \kpc$, so the source is
behind the galactic center.  We assume the source is a member of the
bulge, so we take $f_S$ to be the above velocity distribution function
of bulge stars, and then we evaluate Eq.~\ref{eq-like} separately for
$\rho_L, f_L$ appropriate for lenses either in the disk or the bulge.

Figure~\ref{fig-like} shows the result of Equation~\ref{eq-like} as a
function of lens distance, for the observed $\vhat = 193 \kms$.  The
lower solid line shows disk lenses only, and the upper solid line
shows the sum of disk + bulge lenses; the relative areas indicate that
there is about an $80\%$ probability that this lens belongs to the
bulge.
 
We define a median distance $x_{\rm med}$ such that half of the
integrated likelihood arises from $x < x_{\rm med}$, giving $x_{\rm
  med} = 0.77$ and a median mass estimate of $M_{\rm med} = 0.67
\Msun$.  Similarly, we define a 90\% confidence interval $(x_1,x_2)$
such that 5\% of the integrated likelihood arises from each of $x <
x_1$ and $x > x_2$; the resulting interval is $0.52 < x < 0.943$,
which translates into a mass interval of $0.21$ to $3.2 \Msun$.  These
results are relatively insensitive to the details of the galactic
model, since the constraints on $x$ are dominated by the drop in
$\rho_{Bulge}$ for $x \lesssim 0.6$ and by the geometrical factor
$\sqrt{x(1-x)}$ for $x \rightarrow 1$.
 
For reference, Figure~\ref{fig-likev} shows $\vhat \int {\cal
  L}(x;\vhat) \, dx$, i.e. the lensing rate per unit $\log \vhat$ for
disk and bulge lenses.  The observed value of $\vhat = 193 \kms$ is
well within the range of expected values.  The areas under the two
curves reflect the fact that bulge lenses produce about 3 times the
event rate of disk lenses for the above models.  This figure shows
that there is a large overlap in the distributions; events with $\vhat
\lesssim 150 \kms$ are produced almost entirely by bulge lenses, but
the more common events with larger $\vhat$ arise from both disk and
bulge lenses; thus, proper motion measurements towards the bulge are
not as useful as parallax measurements for constraining the lens
distance.  (This has been previously noted by \citeN{han-bar}).

Note that this is distinctly different from the LMC case, where either
proper motion or parallax measurements provide a good separation of
the various lensing populations \cite{gould-rev}.

\subsection{Constraints on Lens Luminosity}

As seen above, if the lens is more distant than $x > 0.9$, it has a
relatively large mass. If it is a hydrogen-burning star, we can
constrain this possibility as follows.  For the above proper motion of
$\omega \approx 4.5$ mas/year, the lens is completely unresolved from
the source during our observations, thus the observed light--curve of
our `object' is simply the sum of that from the source, the lens and
possibly other superposed stars.

Any flux from the lens would add a constant un--magnified baseline
which would distort the microlensing fit.  Although in most cases this
additional flux would be a small fraction of the flux of the source
star, with the high precision measurements here this would be
detectable even if the lens is considerably fainter than the source.

We have fit ``blended'' microlensing to data of magnification $A <
10$.  This fit is of the form $f_i(t) = A(t) f_{0i} + f_{Ui}$, for
$i=$ each passband, where $f_{0i}$ is the baseline flux of the source
star and $f_{Ui}$ is the total flux of unlensed stars superposed on
the source, i.e. the sum of flux from the lens and any other
superposed stars.

We note that the $\chi^2_{dof}$ of all light--curves in these blended
fits are similar to those in the extended source fits.  We therefore
exclude the MACHO data in our lens brightness estimate due to
excessive baseline scatter; however, this turns out to be unimportant.
The most follow--up data with $A < 10$ are contained in the CTIO and
UTSO observations.  For the CTIO R, UTSO R, and UTSO V passbands, we
find $f_{Ui} / f_{0i} \approx (-0.021 \pm 0.016), (-0.019 \pm 0.017)$,
and $(-0.015 \pm 0.016)$, respectively, consistent with zero flux from
the lens.  For completeness, we note that similar results are obtained
from the MACHO data, with $f_{Ui} / f_{0i} \approx (-0.012 \pm 0.017)$
and $(-0.026 \pm 0.018)$ for MACHO V and MACHO R, respectively.  MJUO
and Wise data did not contain enough baseline observations to
determine accurately the amount of unlensed flux.  In the following, we take
$0 \pm 0.02$ as a conservative limit.

If the lens is a main-sequence star, we may predict its apparent
brightness using the mass-distance relation of Eq.~\ref{eq-mx}.
Assuming a main-sequence V-band mass-luminosity relation $L / L_\odot
= (M / M_\odot)^{3.5}$, and $1.5$ magnitudes of extinction, we find that
the apparent brightness of a main-sequence lens would be e.g. $1\%$ of
that of the source star for $x = 0.83$, and $10\%$ for $x = 0.90$.
Note that the implied lens brightness {\em increases} with distance
since the rapid rise in $M(x)$ outweighs the $r^{-2}$ term; note also
that a giant lens is excluded at any distance.

Instead of applying a sharp limit $x < 0.88$, it is more rigorous to
multiply the likelihood function of Eq.~\ref{eq-like} by the Gaussian
probability that the lens brightness is consistent with the above
constraint on $f_U/f_0$.  This causes a rapid roll-off in the
likelihood function for $D_L > 7.5 \kpc$, as shown by the dotted
curves in Figure~\ref{fig-like}.
Assuming a main-sequence lens, we can re-compute the constraints on
$x$ and $M$ from the likelihood function with the brightness factor,
giving a median $x = 0.73$, a mass of $M_{med} = 0.53 \Msun$, and 90\%
confidence intervals of $0.48 < x < 0.84$ and $0.18 < M < 1.05 \Msun$.
Thus, the inclusion of the lens brightness constraint reduces the
median mass only slightly, but considerably strengthens the upper
limit.

\section{Spectral Variation during MACHO 95--30}
\label{sec-spectra}

In addition to the CTIO observations discussed in
Section~\ref{sec-spec}, we have obtained spectra at Mount Stromlo and
Keck~\footnote[5]{The Keck telescope project was made possible by a
generous grant from the W.M. Keck Foundation.}  Observatories.
Table~\ref{tab-spec} lists the complete catalogue of spectroscopic
observations of this event. Figure~\ref{fig-specdate} schematically
shows the location of the lens with respect to the source when these
different observations were made.  Nightly spectra of the source star
were taken between August 18 and 25, 1995 (from 2.8 to 9.8 days after
peak magnification), with the Cassegrain spectrograph at the MSO
74--inch telescope.  They cover the wavelength interval 6240--6770 \AA
, with a dispersion of 0.90 \AA\ per pixel, and a resolution FWHM =
4.6 \AA\ as determined from the comparison lamp spectra. The total
exposures were typically 1000 seconds long. These MSO spectra are
wavelength calibrated and sky subtracted, but not fluxed. The S/N per
resolution element is listed in Table~\ref{tab-spec}, along with other
relevant data.

The CTIO and MSO spectra have similar resolution, with the exception
of the September 27 CTIO spectrum, which has lower resolution.  We
rebinned the CTIO and MSO spectra in order to compare them directly.
Figure~\ref{fig-specall} shows the spectral sequence, which
constitutes one of the most extensive and homogeneous series of
spectroscopic observations of a microlensing event to date (see also
\citeNP{benetti}).  It is clear that the microlensing does not change
the spectral type of the star, nor does it strongly affect the major
spectral features.  However, subtle effects may appear in some
spectral lines, as discussed by \citeN{loeb-sass}, which warrants a
more careful comparison among the spectra.

Before this discussion, we note that the MSO spectrum of August 18
(the first spectra taken {\it after} the lens transit) is anomalous,
showing a dip at $\lambda 6520$\AA\ that looks like an unidentified
bandhead.  Although we have examined several possible sources for this
feature, including checking the flat fields and comparing the spectra
of other stars in the field, the cause of this dip is still
unexplained.

Each spectrum was divided by the median combination of all spectra
(this operation was repeated before and after continuum subtraction,
in order to account for possible differences due to flux calibration
of the CTIO spectra but not of the MSO ones).
Figure~\ref{fig-specmed} shows the sequence of these ratios. While
there are no strong deviations, the two spectra taken closest to
maximum magnification show stronger H$\alpha$ than the rest. Also,
these two spectra show stronger TiO bands than the following ones. The
total TiO absorption, however, started to climb steadily after 18
August, 1995.  We measured the equivalent widths of several spectral
features of interest within 6240--6770 \AA\ with the SPLOT package
within IRAF.  Figure~\ref{fig-ew} shows the equivalent width of
H$\alpha$, and the combined intensity of 4 TiO bands in the 6650--6750
\AA\ region.  Included are conservative error bars based upon
uncertainty in placing the local continuum.  The equivalent width of
the H$\alpha$ line was scaled to the equivalent width measured in the
HIRES spectra of 18 and 19 August, 1995. These equivalent widths were
measured with respect to the local pseudo continuum, in the same way
for all the spectra, in order to avoid systematic effects.

While we cannot explain in detail the behavior of these spectral
features, changes in the equivalent width of these lines have been
predicted by \citeN{loeb-sass}.  These changes are due to limb
brightening effects in the cores of resonance lines (like H$\alpha$)
due to a very extended photosphere.  Resonant line scattering may also
affect the optical TiO bands in the case of M giants like the MACHO
95--30 source \cite{sasselov}.  Detailed modeling of the spectra of
this event in particular is needed, as this will help with the
interpretation of forthcoming similar events.

Three spectra were also taken in the nights of August 18, August 19,
and August 21, about 5 hours later than the MSO spectra, using HIRES
at the Keck 10--m telescope. These high resolution spectra also have
high S/N, as listed in Table~\ref{tab-spec}, which includes the S/N
per resolution element of the lowest and highest orders. The spectra
of August 18, August 19, and August 21 consist of 30, 25, and 40
orders, respectively. The blue region below about 4000\AA\ covered in
the August 21 spectrum does not have much information due to low
counts.  We have checked different resonance lines (CaII, H$\alpha$,
H$\beta$, H$\gamma$, etc), finding no large variations.  The August 21
spectrum has coverage extending to the blue, including the Ca H and K
lines, but misses the H$\alpha$ line.  The spectra of August 18 and 19
do not include these Ca lines.  The H$\alpha$ equivalent width differs
by only 4\% between August 18 and 19, as measured from the high
quality HIRES spectra (1.087 vs 1.039\AA).  This variation occurs in
the core of the line.  Otherwise, there is no significant change
between the H$\alpha$ profiles of August 18 and 19.  The strong TiO
band heads at $\lambda\lambda$ 4954, 5166, and 5447 \AA\ are present
in all three HIRES spectra. The difference of about 10\% in strength
of these bands between August 18 and 21 seen in the MSO spectra is
confirmed.

We have also fit the NaI (D) line profiles in the HIRES spectra with
an interstellar cloud model which uses 12 clouds over the velocity
range $-60.8 \kms$ to $+202.3 \kms$. The clouds with extreme
velocities only fit one line of the Na doublet because of problems
with the continuum determination and blending with other lines.  The
interstellar clouds indicate that the star is at a great distance and
therefore it is certainly a giant, as discussed in
Section~\ref{sec-spec}.

We also measured the radial velocities of the MSO spectra using the
cross-correlation routine FXCOR within IRAF.  From the MSO spectra, we
measure a mean velocity of $V_r = -76 \pm 4 \kms$, which is in
excellent agreement with the mean velocity of $V_r = -80 \pm 5 \kms$
from the CTIO spectra, reduced and calibrated independently
(Section~\ref{sec-spec}).  The velocities of the eight MSO spectra
agree with each other within an rms of $12 \kms$.  We measured also
the radial velocity of the August 21 HIRES spectrum, $V_r = -86.2 \pm
3.9 \kms$, using the strong Cr lines at $\lambda\lambda$ 4254, 4274,
and 4289 \AA.  All these velocities are in excellent agreement, and we
adopt a final value of $V_r = -81 \pm 5 \kms$ for the source star in MACHO
95--30.

Subtle radial velocity variations (of the order of $1-2 \kms$) may
also be expected during microlensing events such as this one
\cite{maoz-gould}, and can give an independent measurement of the
proper motion.  Alternatively, combining the radial velocity variation
with the extended source magnification effect, one can determine the
projected rotation velocity of the star \cite{gould-rot}.  We note
that detecting such velocity differences may be possible using the
HIRES spectra, after detailed corrections and modeling that are beyond
the scope of the present paper.

In summary, we have obtained a large number of spectra of the
microlensed source star during MACHO Alert 95--30.  These spectra show
subtle variations, which support the predictions of existing
microlensing and stellar models.  These spectra demonstrate that
differential magnification of an extended source star during a
gravitational microlensing event may be used to probe otherwise
inaccessible stellar atmosphere ``fine structure''.

\section{Conclusions}
\label{sec-conclusions}

We have observed the breakdown of the point--source approximation for
the first time in a ``normal'', single lens microlensing event.  This
deviation was anticipated based upon real--time information provided
by MACHO--GMAN, and dynamically scheduled for GMAN follow--up
observations throughout the event.  In effect, we have ``resolved'' a
star which is $\sim 9\kpc$ from Earth.  The resulting deviation from
point--source microlensing provides a second constraint equation for
the system, partially reducing the ambiguity between lens mass,
position, and transverse velocity.

Measurement of the effective radius of the source star during the
microlensing event provides an estimate of the scale of the lens's
Einstein radius.  With additional information about the radius and
distance of the source, we determine the angular velocity of the lens
with respect to our line of sight to the source.  This allows a
statistical estimate of the lens mass by assuming velocity
distributions of sources and lenses.  We conclude that this event is
due to lensing by an object of mass $0.67^{+2.53}_{-0.46} \msun$ (at
the 90 \% confidence level), such as a white dwarf or a neutron star,
which is most likely in the bulge (80 \% c.l.).  If the lens is a
main--sequence star, the upper end of this mass range is excluded, and
our mass estimate becomes $0.53^{+0.52}_{-0.35} \msun$.  The results
of this paper are attributable to the coordinated efforts of GMAN
observers, and are a robust endorsement for microlensing follow--up
implemented at 1--m class telescopes.

We have also detected for the first time subtle variations in the
spectra of a star undergoing microlensing.  Anomalously strong
equivalent widths of H$\alpha$ and TiO were detected while the lens
was in transit across the source, an effect anticipated due to varying
line widths across the face of the star.  Given our ability to predict
limb crossings in real--time, this technique may be used to map out
the center--to--limb variations in the spectrum of a future
microlensed source.

\acknowledgements
\section*{Acknowledgments}
We would like to thank Antonio Claret for providing limb--darkening
coefficients for the MACHO passbands.  The authors also wish to thank
the skilled GMAN observers at UTSO, Boyd Duffee and Felipe McAuliffe.
We are very grateful for the skilled support of the staff and
observers at CTIO, in particular Bob Schommer, for flexibility in
scheduling GMAN follow--up observations.  We thank the NOAO for making
nightly use of the CTIO 0.9~m telescope possible.

Astronomy at Wise Observatory is supported by grants from the Israel
Science Foundation.

The MOA group thanks the Directors of MJUO and KEK for support, and
Messrs G. Nankivell and N. Rumsey and staff members of Perth
Observatory, MJUO, Camex Ltd and the Auckland University Science
Faculty Workshop for assistance.  Grants by Auckland and Victoria
Universities, NZ/Japan Foundation, Ministry of Education, Science,
Sports and Culture of Japan, KEK Laboratory, NZ Lottery Grants Board,
NZ Ministry of Research, Science and Technology and the NZ Marsden
Fund are gratefully acknowledged.

Work performed at LLNL is supported by the DOE under contract
W7405-ENG-48.  Work performed by the Center for Particle Astrophysics
personnel is supported in part by the Office of Science and Technology
Centers of NSF under cooperative agreement AST-8809616.  Work
performed at MSSSO is supported by the Bilateral Science and
Technology Program of the Australian Department of Industry,
Technology and Regional Development.  WJS is supported by a PPARC
Advanced Fellowship.  KG acknowledges support from DOE Outstanding
Junior Investigator, Alfred P. Sloan, and Cottrell awards.  CWS thanks
the Sloan, Packard and Seaver Foundations for their generous support.

\clearpage

\figcaption[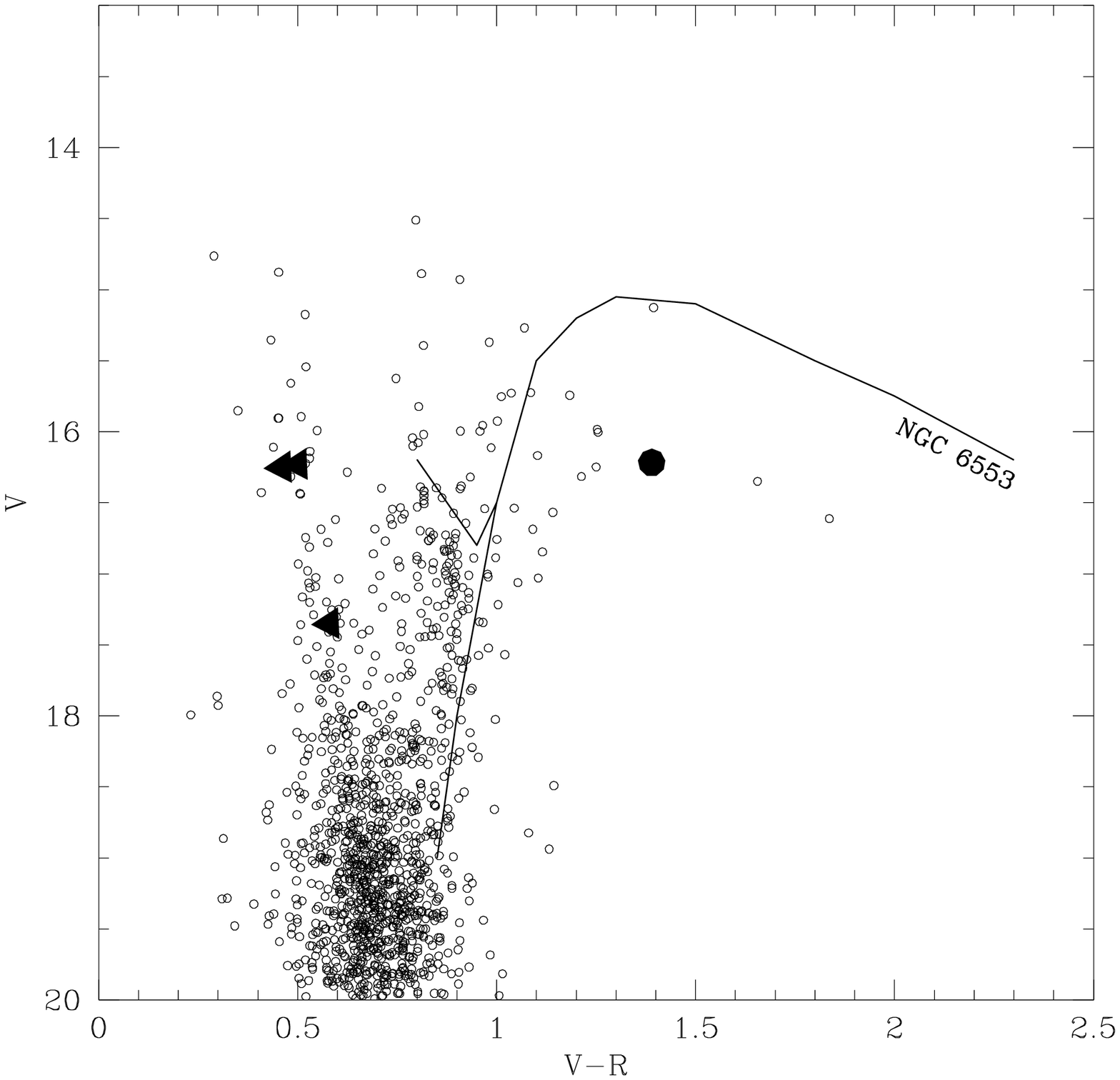]{\label{fig-cmd}
  Single epoch optical $V$ vs $V-R$ color-magnitude diagram of the
  field surrounding the source star, which is indicated with a filled
  circle.  Three RR Lyr used to estimate the reddening are plotted as
  triangles.  The fiducial loci of the giant branch of the metal--rich
  globular cluster NGC~6553 ($[Fe/H] = -0.2$) from Ortolani et al.
  (1990) is also indicated.  }

\figcaption[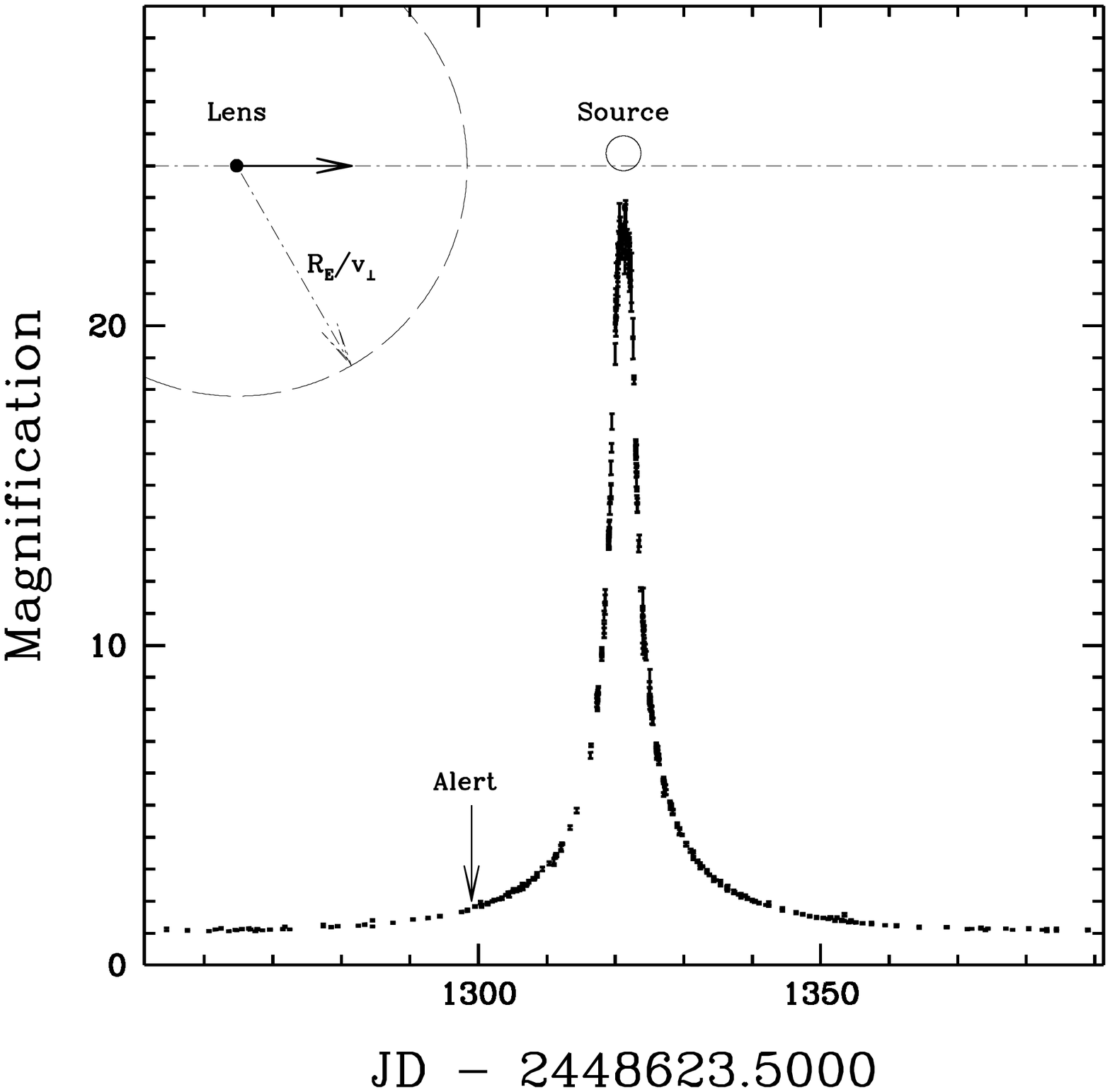]{\label{fig-all} 1995 light--curve of MACHO Object
  101.21821.128 (Alert 95--30).  The data represent MACHO--GMAN
  observations of this event in all passbands within a window of 70 days
  around peak magnification.  The date this event was detected by the
  MACHO Alert system is indicated with an arrow.  An additional
  schematic relates the scale of the lens's Einstein radius to the
  angular size of the source star, and indicates transit of the lens
  across the source face.}

\figcaption[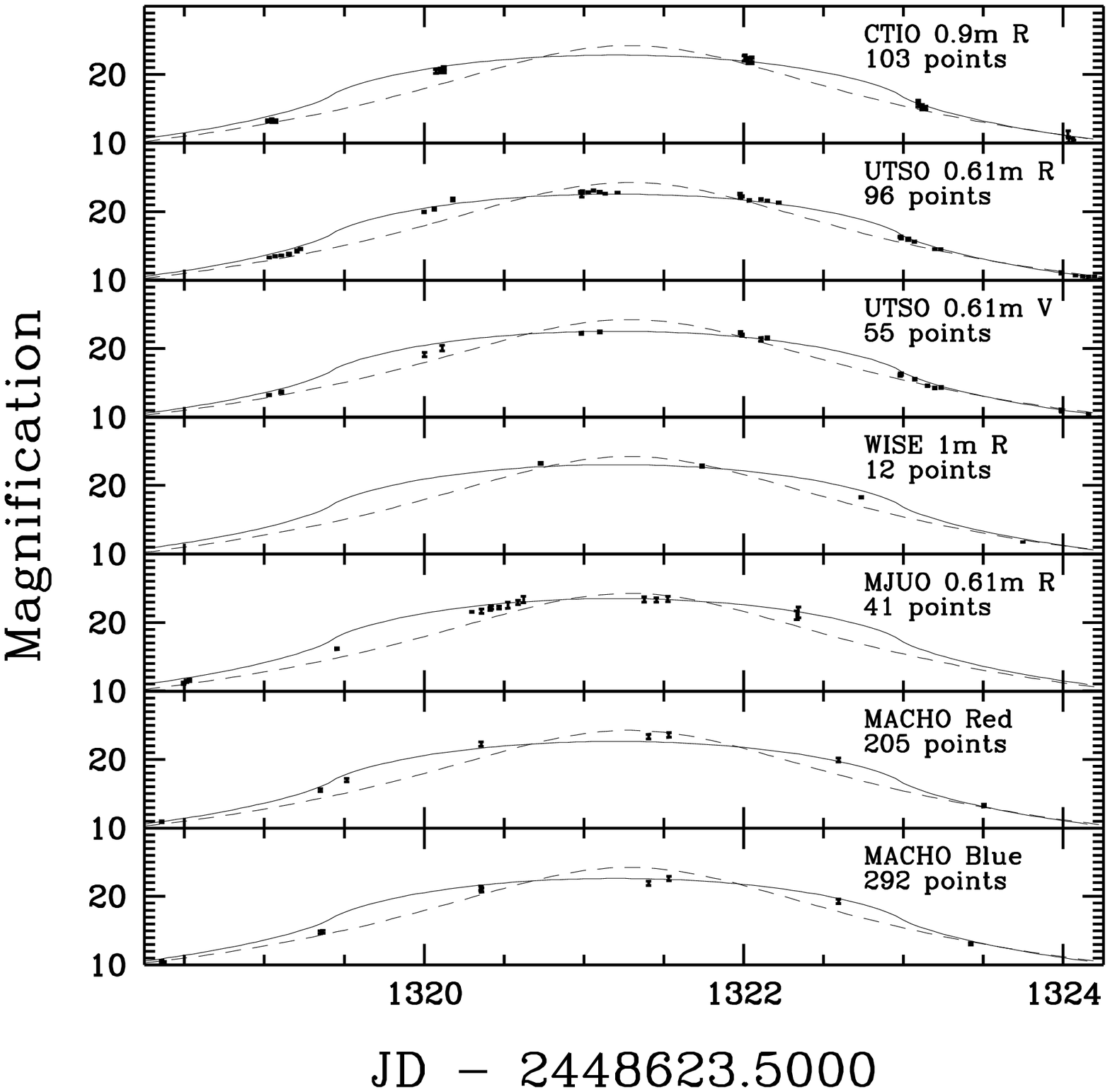]{\label{fig-peak} Peak of MACHO Alert 95--30, showing the best standard
  microlensing fit to the data (~-~-~-~) and an extended source
  microlensing fit incorporating source limb--darkening (------).}

\figcaption[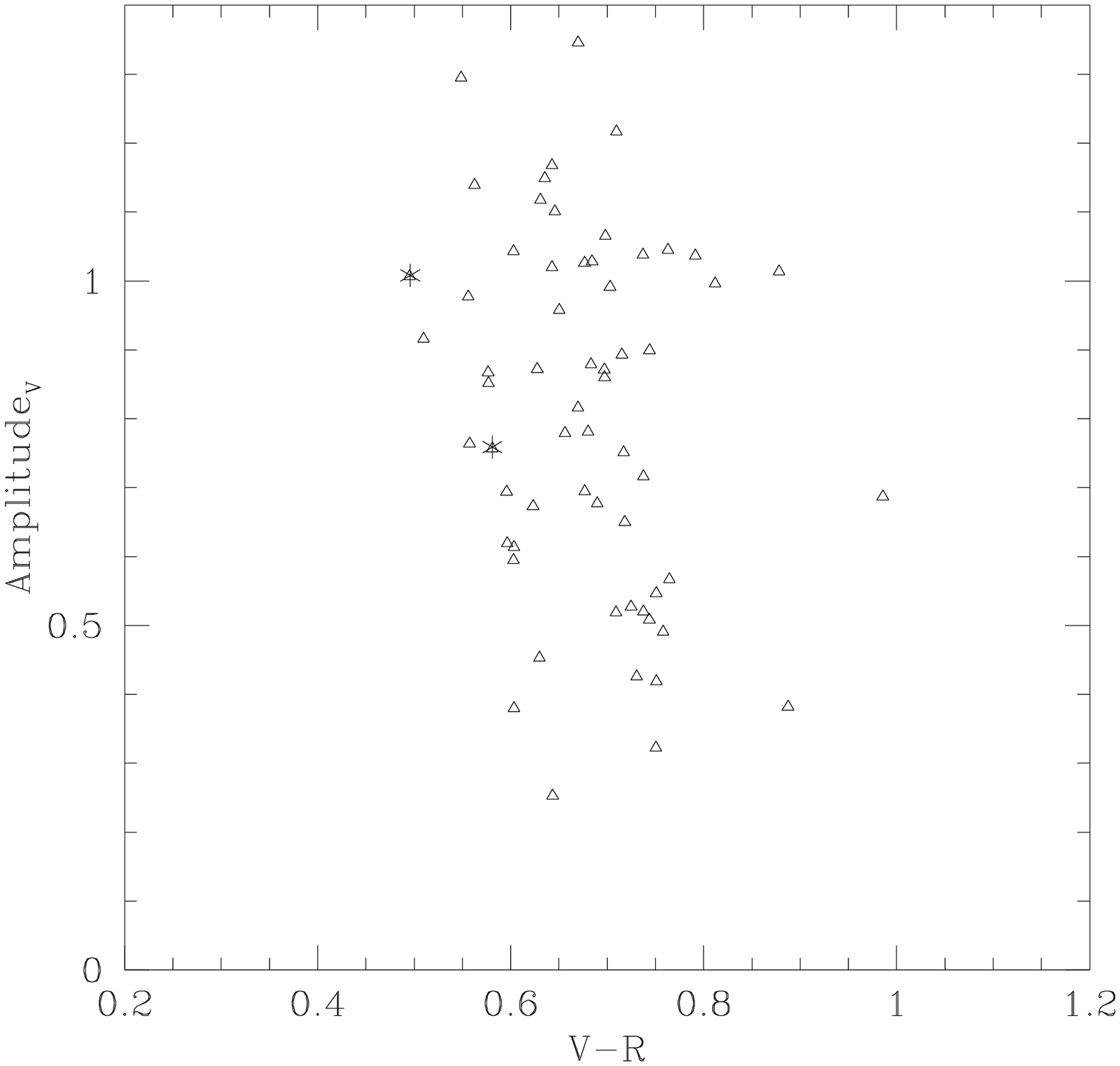]{\label{fig-rrl}
  Amplitude vs. color of the RR Lyr type ab in the MACHO field where
  the event is located. The two RRab closest to the source star are
  indicated with asterisks.  We find $A_{V} = 1.35$ mag.  }

\figcaption[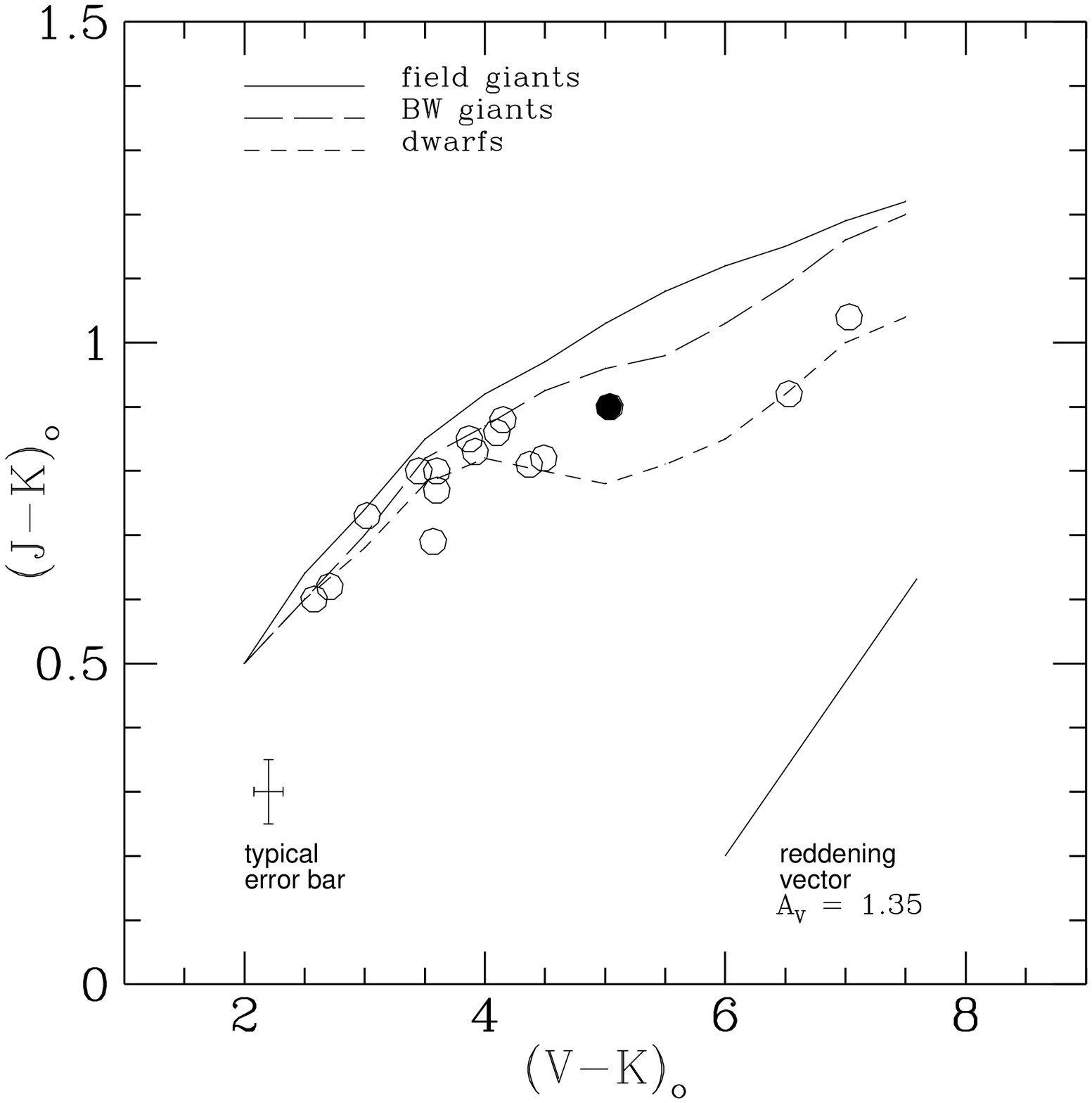]{\label{fig-ccd}
  Optical-infrared $(V-K)_{0}$ vs $(J-K)_{0}$ color-color diagram for
  the giants brighter than 1 mag above the horizontal branch in the
  field surrounding the source star.  The fiducial loci of field giants,
  dwarfs, and Baade's Window giants from Frogel \& Whitford (1987) are
  indicated with the solid, short-dash, and long-dash lines,
  respectively.  }

\figcaption[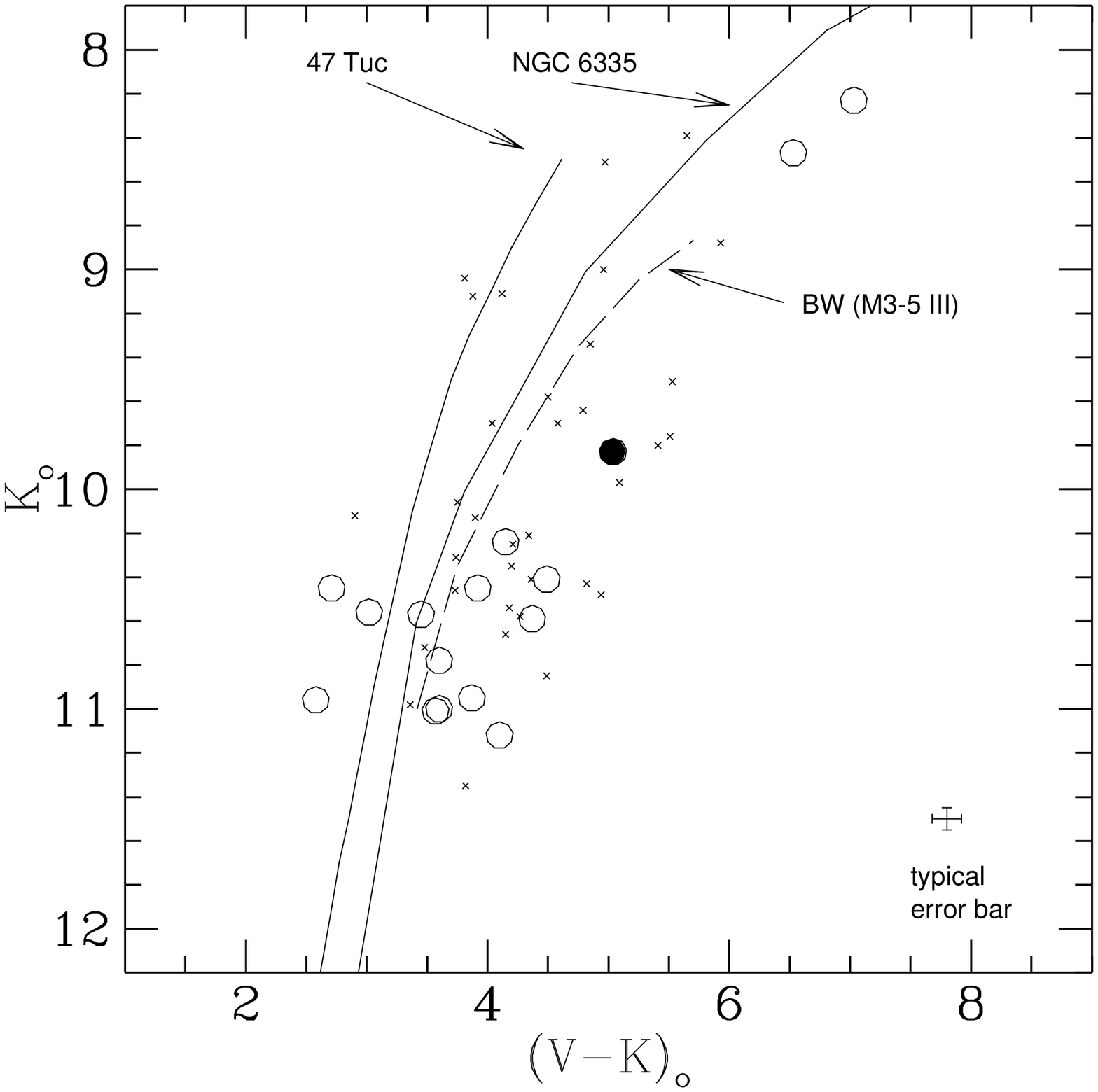]{\label{fig-oircmd}
  Optical-IR $K_{0}$ vs $(V-K)_{0}$ color-magnitude diagram of the
  field surrounding the source star.  The fiducial loci of the giant
  branches of the metal--rich globular clusters 47~Tuc ($[Fe/H] =
  -0.7$), and NGC~6553 ($[Fe/H] = -0.2$) from Guarnieri et al. (1996)
  are indicated.  Additionally, we have plotted BW M3--5 giants
  (Frogel \& Whitford, 1987) as small crosses and drawn our own
  estimated fiducial line through these points.  }

\figcaption[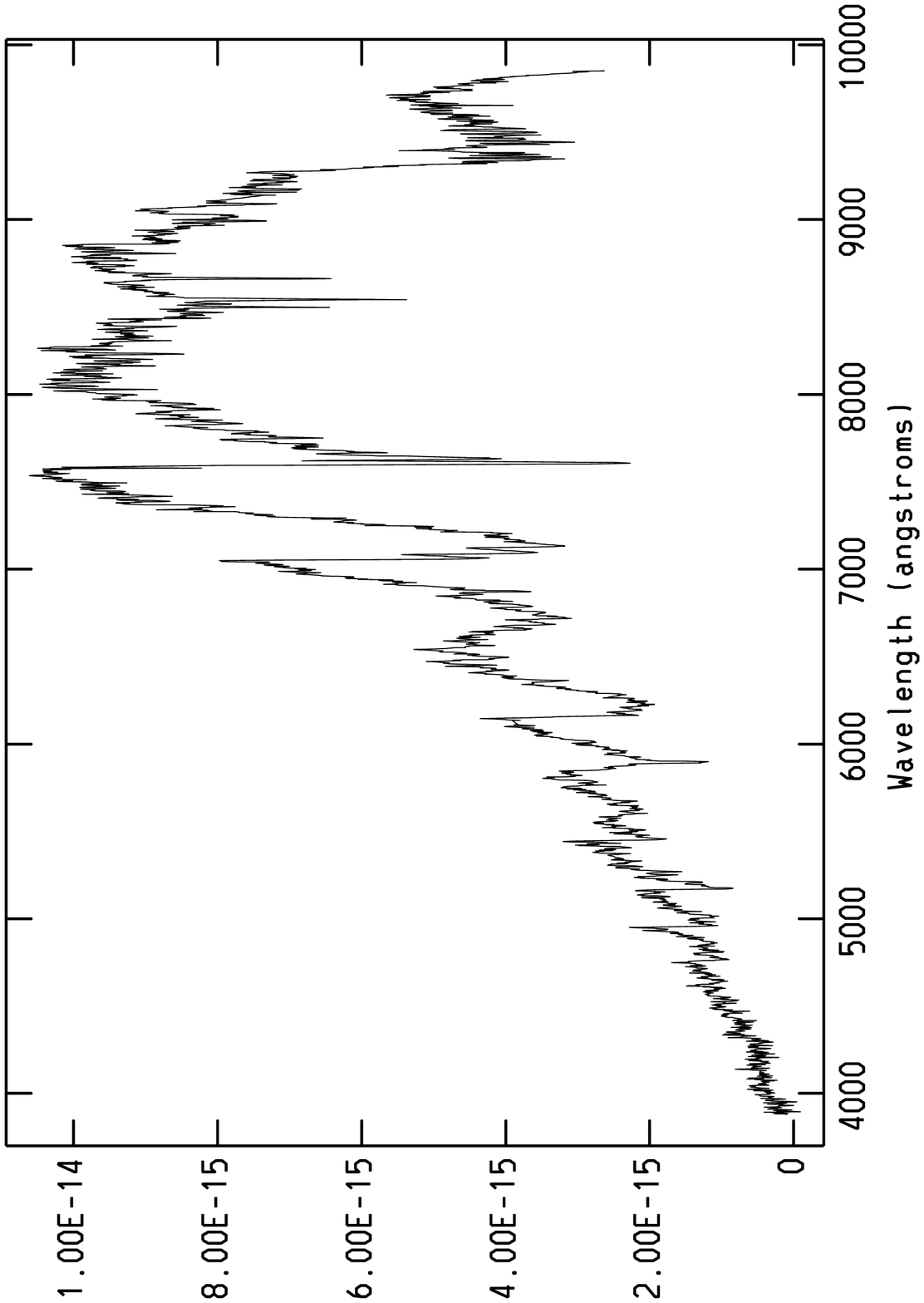]{\label{fig-spec1}
  Baseline spectrum of the source star obtained at CTIO on September
  27, 1995, 42 days after maximum magnification. The flux calibration
  is unreliable beyond 9000\AA. }

\figcaption[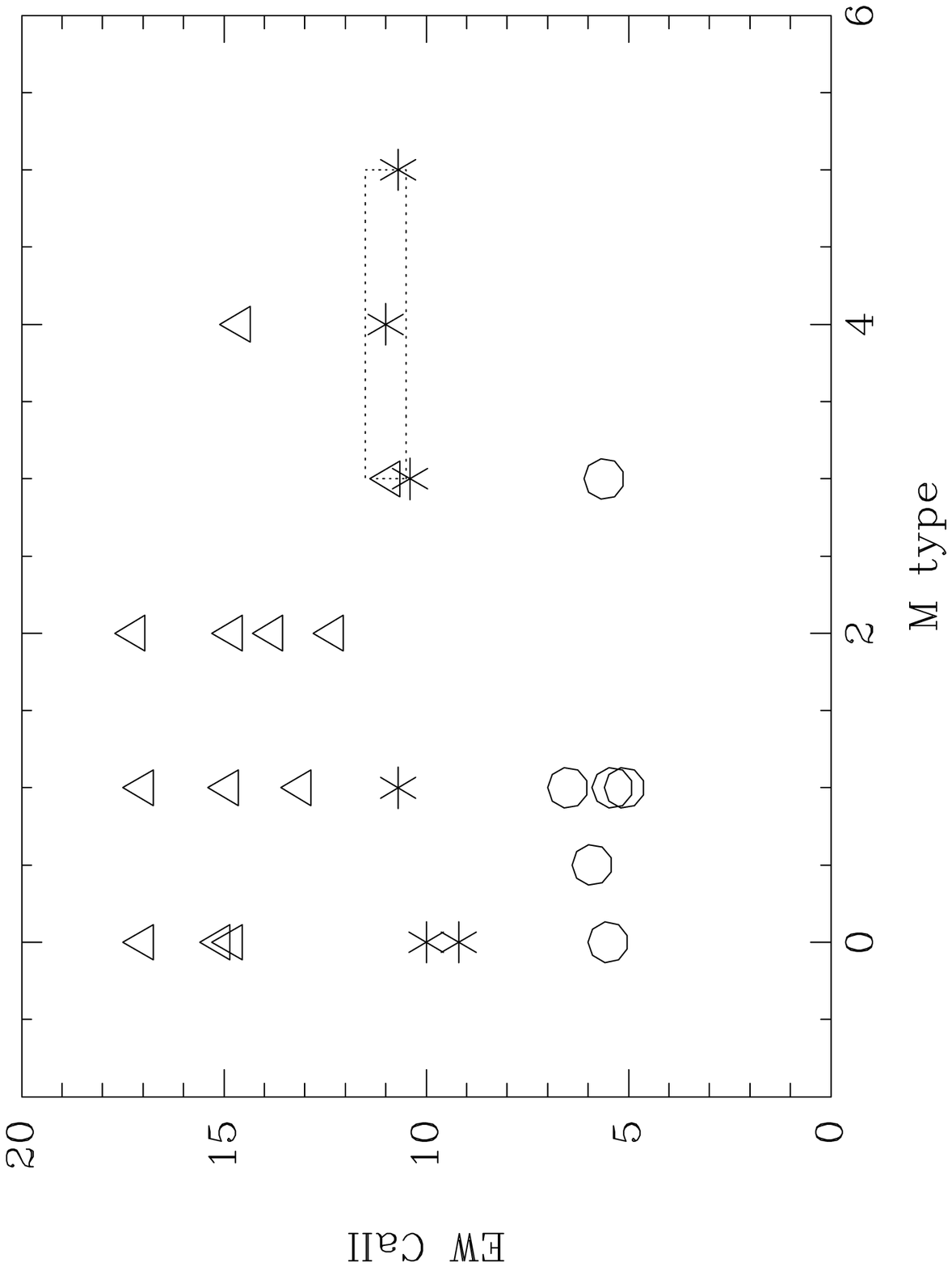]{\label{fig-ew1}
  CaII equivalent width in \AA\ $vs$ M spectral sub-type for dwarfs
  (circles), giants (asterisks), and supergiants (triangles). The box
  encloses the $\pm 1 \sigma$ measurements of the source star in MACHO
  95--30. }

\figcaption[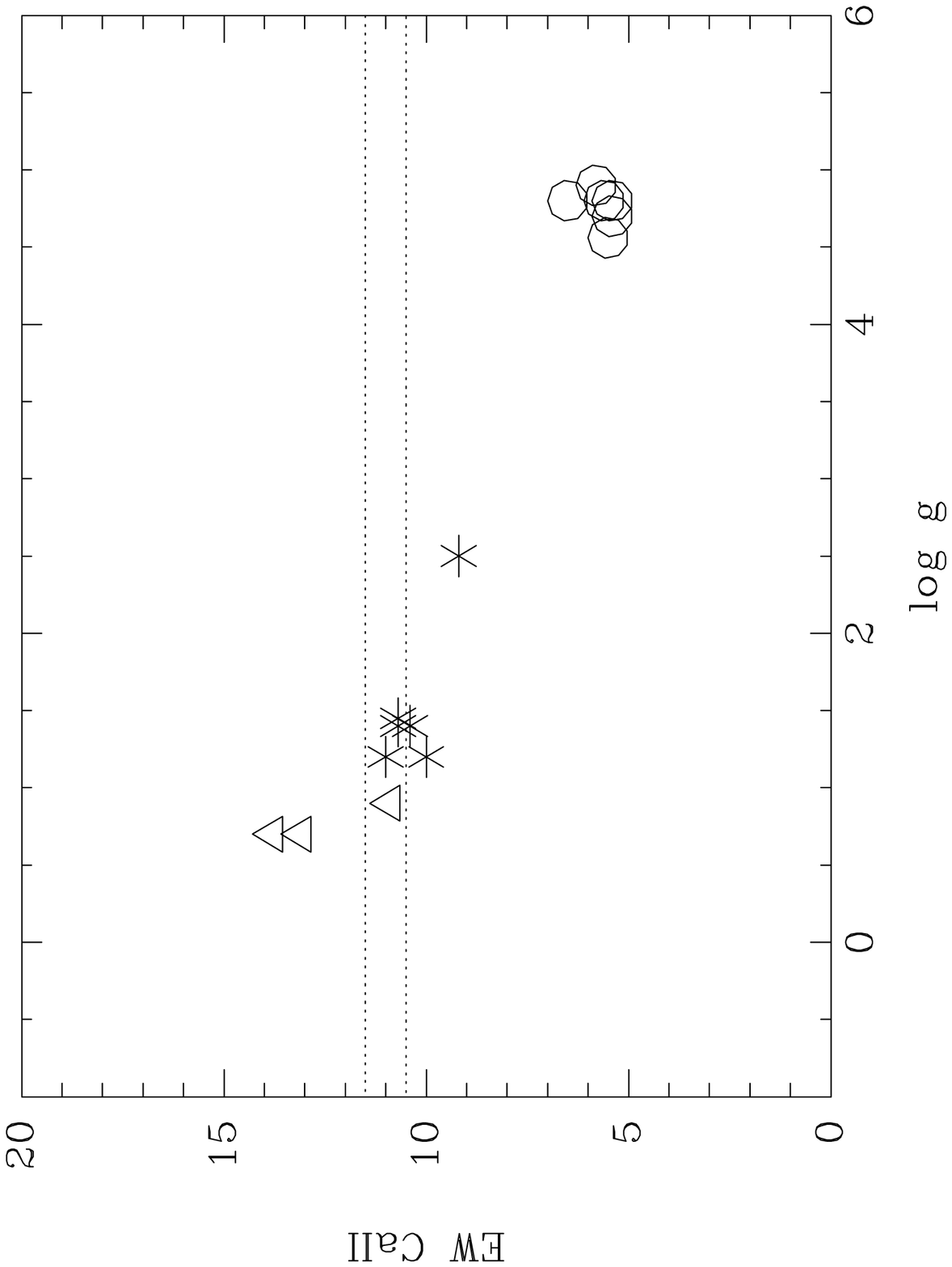]{\label{fig-ew2}
  CaII equivalent width in \AA\ $vs$ log $g$, with the dotted lines
  indicating the CaII equivalent width of the source star in MACHO
  95--30 ($\pm 1 \sigma$). }

\figcaption[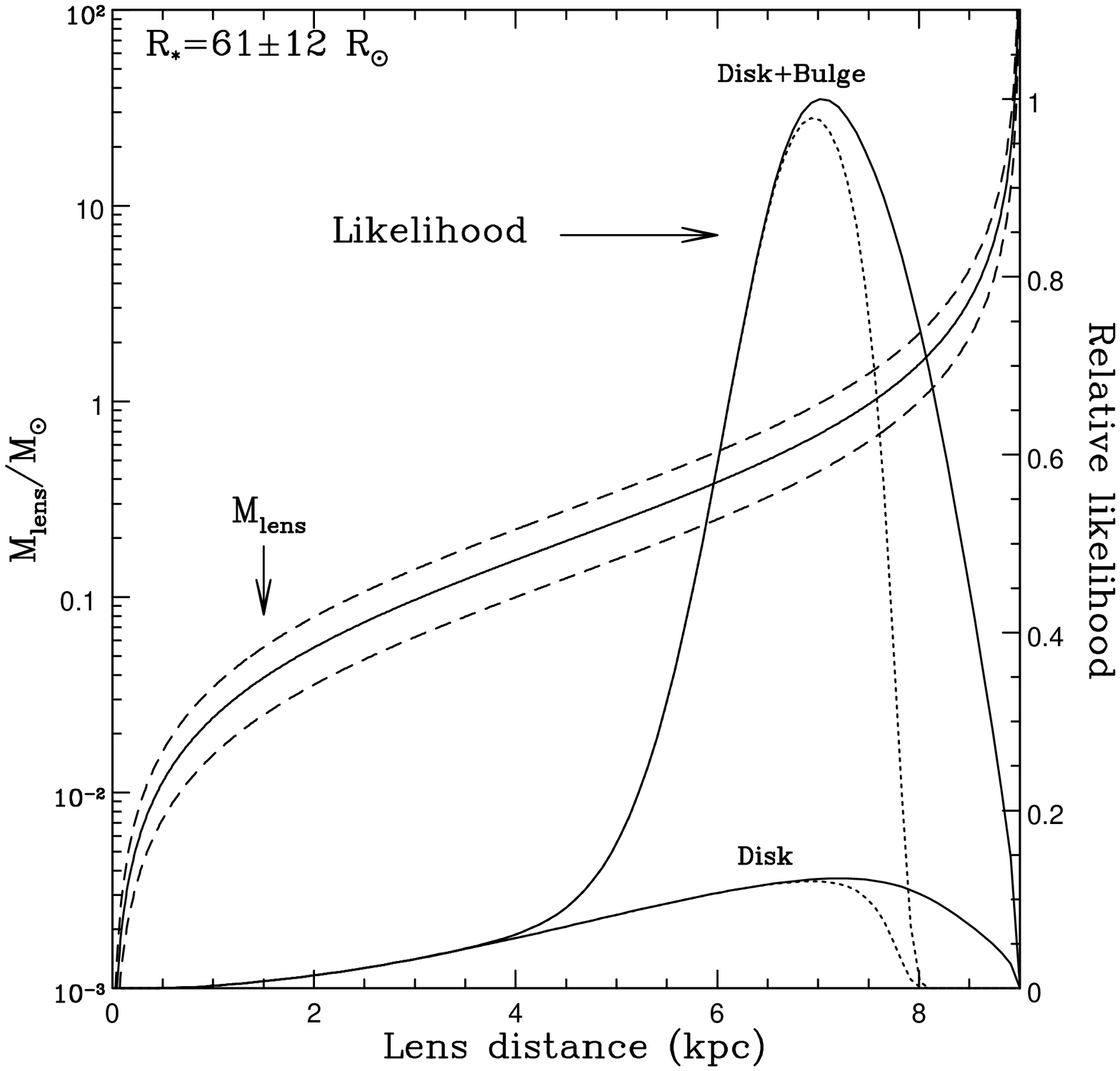]{\label{fig-like} 
  Lens mass plotted as a function of its distance, using
  Equation~\ref{eq-mx}.  This assumes a $61 \rsun$ source, with error
  contours provided at $\pm 12 \rsun$, and event parameters listed in
  Table~\ref{tab-stat2} (fit 3).  The solid likelihood curves show
  relative probabilities for disk and bulge lenses as a function of
  their distance.  The dotted lines include an upper limit on the
  brightness of a main--sequence lens.  }

\figcaption[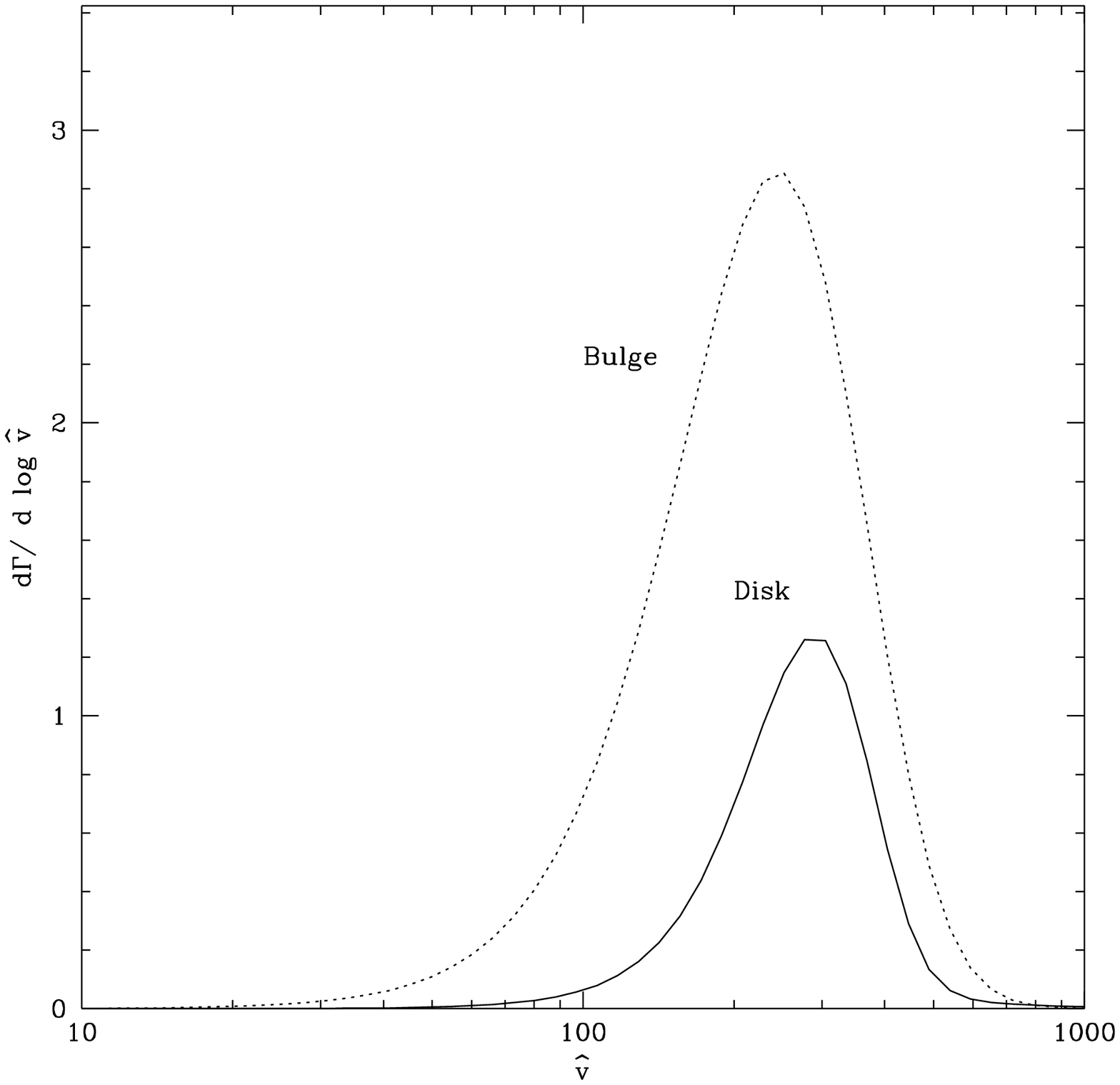]{\label{fig-likev} 
  Lensing rates per unit log $\vhat$ for disk and bulge lenses,
  indicated with the solid and dashed lines, respectively. }

\figcaption[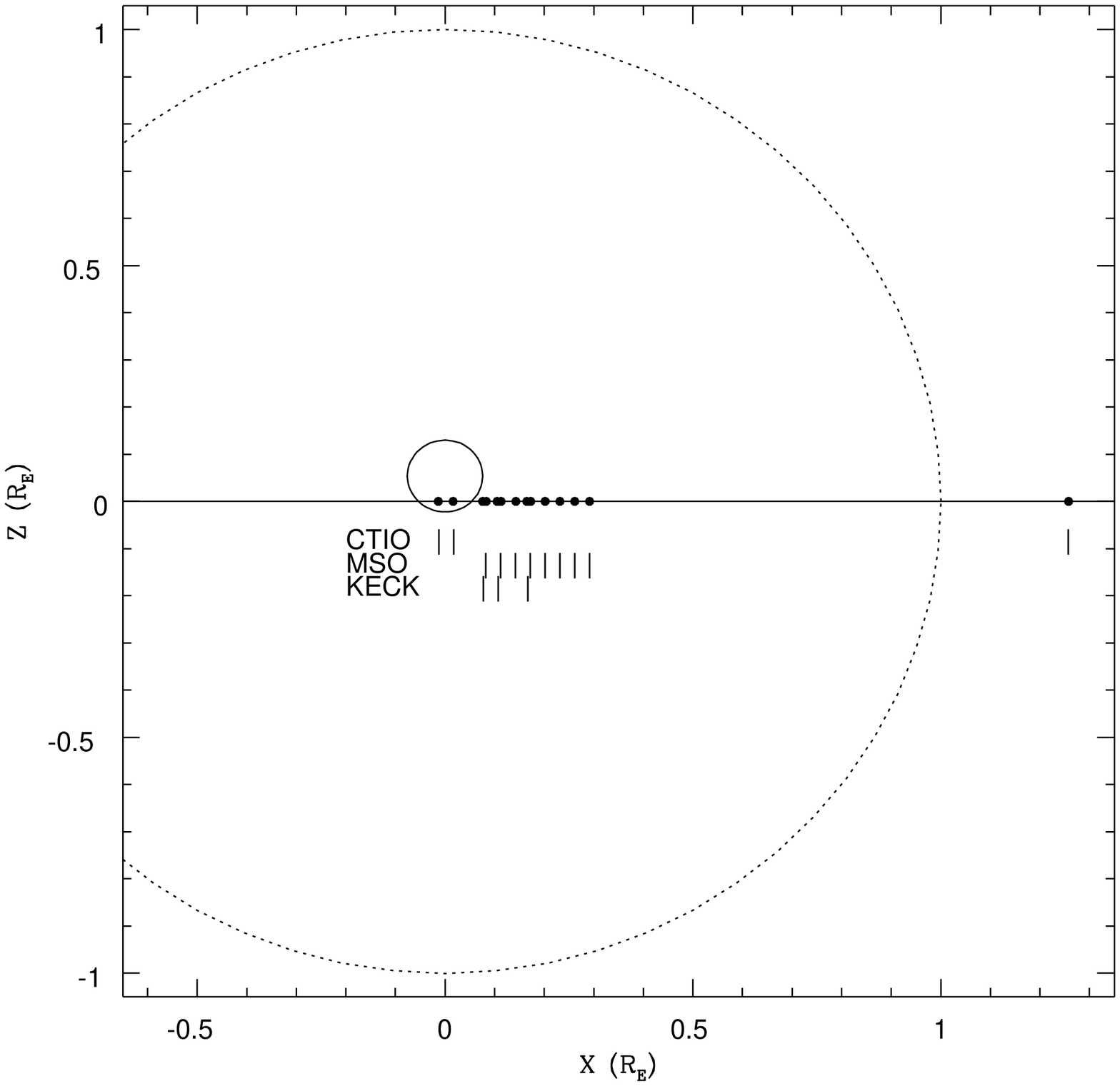]{\label{fig-specdate}
  Spectral observations listed in Table ~\ref{tab-spec} plotted on the
  source plane. The relative sizes of the star (solid circle) and the
  lens's Einstein radius (dotted circle) are plotted to scale, in
  units of $\Re$.  The solid points show the position of the lens at
  the different times when spectra were taken. The observatories where
  the spectra were taken are indicated with the tick marks. }

\figcaption[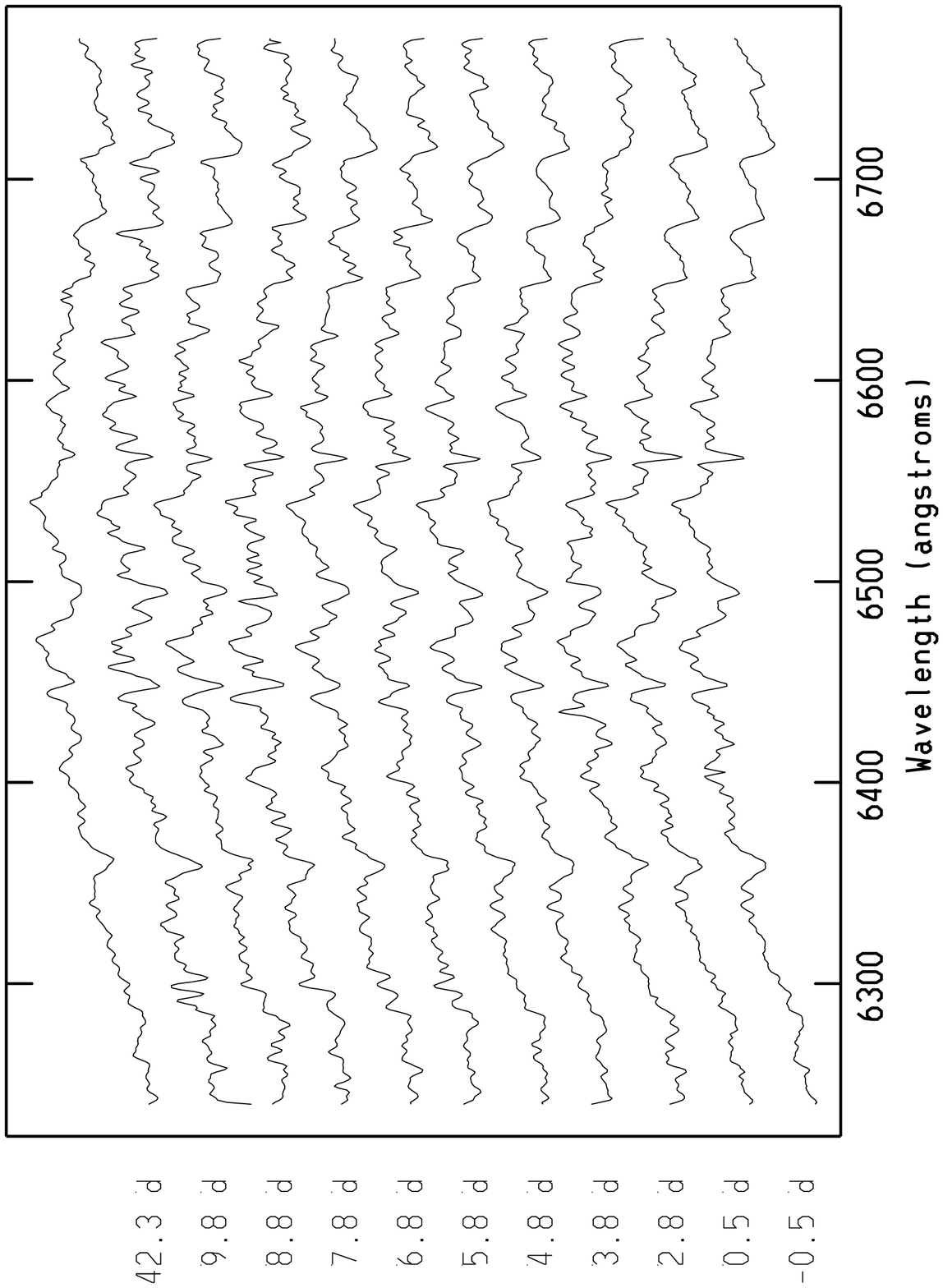]{\label{fig-specall}
  Spectra of the source star taken at CTIO and MSO. All of these have
  similar resolution, except for the top one. Note H$\alpha$ at
  $\lambda 6553$ \AA, and the TiO band-heads at $\lambda \lambda$
  6647, 6676, 6711, and 6742 \AA.  Days from the peak of the
  microlensing event are indicated along the vertical axis. }

\figcaption[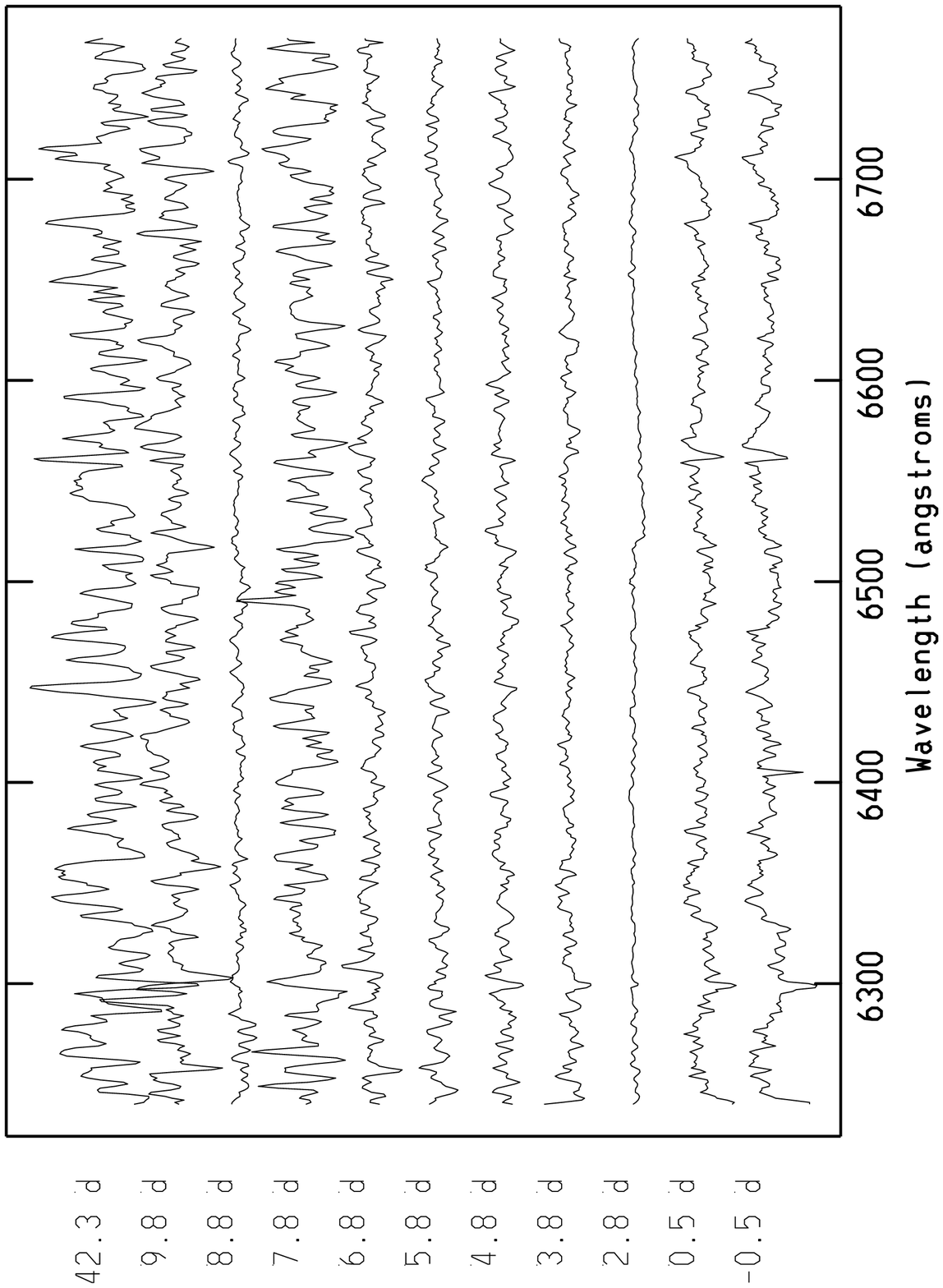]{\label{fig-specmed}
  Same as Figure~\ref{fig-specall}, showing the spectra divided by the
  median combination of all. These have been displaced in the vertical
  direction and arbitrarily scaled, such that the flattest spectra
  here actually display the largest range of excess from the median
  combination.  Note the changes in H$\alpha$ and the TiO bands in the
  two spectra nearest peak magnification. }

\figcaption[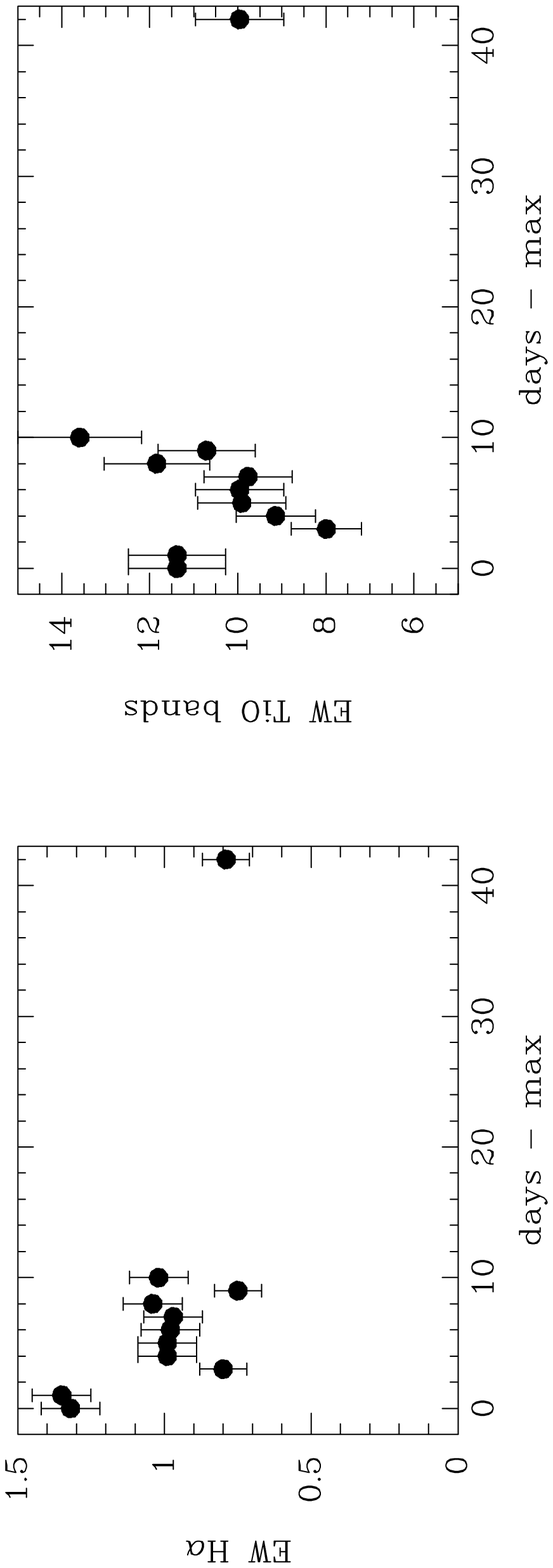]{\label{fig-ew}
  Temporal evolution of the equivalent width in \AA\ of H$\alpha$ and
  the TiO band-heads, measured with respect to the local
  pseudo-continuum. The EW plotted in the right panel are the sum of
  the TiO bands at $\lambda \lambda$ 6647, 6676, 6711, and 6742 \AA. }

\onecolumn


\clearpage
\begin{deluxetable}{llllc}
\small
\footnotesize
\scriptsize
\tablecaption{Global Microlensing Alert Network \label{GMAN} }
\tablewidth{0pt}
\tablehead{
        \colhead {Observatory} &
        \colhead {Latitude, Longitude} &
        \colhead {Aperture} &
        \colhead {Detector} &
        \colhead {Plate Scale (''/pixel)}
}
\startdata
MSO \tablenotemark{a} &
   149 $\fdg$ 00.5 $\fm$ E  ,  35 $\fdg$ 19.2 $\fm$ S &
   1.27 m &
   8 x 2048 x 2048 Loral &
   0.63 \nl
CTIO \tablenotemark{b} &
   70 $\fdg$ 48.9 $\fm$ W  ,  30 $\fdg$ 30.9 $\fm$ S & 
   0.9 m & 
   1024 x 1024 Tektronics & 
   0.40 \nl
MJUO \tablenotemark{c} &
   170 $\fdg$ 27.9 $\fm$ E  ,  43 $\fdg$ 59.2 $\fm$ S &
   0.61 m & 
   1536 x 1024 KAF1600 &
   0.23 \nl
UTSO \tablenotemark{d} &
   70 $\fdg$ 42.0 $\fm$ W  ,  29 $\fdg$ 00.5 $\fm$ S & 
   0.61 m & 
   512 x 512 Photometrics & 
   0.45 \nl
WISE \tablenotemark{e} &
   34 $\fdg$ 45.8 $\fm$ E  ,  30 $\fdg$ 35.8 $\fm$ N & 
   1.0 m & 
   1024 x 1024 Tektronics &
   0.70 \nl
\enddata
\tablenotetext{a} { Mount Stromlo Observatory, Canberra, Australia. }
\tablenotetext{b} { Cerro Tololo Inter-American Observatory, Cerro Tololo, Chile. }
\tablenotetext{c} { Mount John University Observatory, Lake Tekapo, New Zealand. }
\tablenotetext{d} { University of Toronto Southern Observatory, Las Campanas, Chile. }
\tablenotetext{e} { Wise Observatory, Mitzpe Ramon, Israel. }
\tablenotetext{} {Summary of observatories participating in GMAN follow--up
  observations.}
\end{deluxetable}

\clearpage
\begin{deluxetable}{llllcc}
\tablecaption{MACHO Alert 95-30 microlensing statistics \label{tab-stat1} }
\tablewidth{0pt}
\tablehead{
        \colhead {Fit \tablenotemark{a} } &
        \colhead {$\t0$ \tablenotemark{b} } &
        \colhead {$\that$ \tablenotemark{c} } &
        \colhead {$\umin$} &
        \colhead {$\ustar$} &
        \colhead {$\omega$ \tablenotemark{d} } 
}
\startdata
1 & 1321.2  (1) 
  & 67.70   (13) 
  & 0.04069 (12) 
  & 0      
  & ??? \nl
2 & 1321.2  (1)
  & 67.36   (1)
  & 0.05579 (1)
  & 0.07335 (1)
  & 22.1    (50) \nl 
3 & 1321.2  (1)
  & 67.28   (27)
  & 0.05408 (20)
  & 0.07561 (9)
  & 21.5    (49) \nl 
\enddata
\tablenotetext{a} { Fit 1 is the best standard microlensing fit to the data,
  Fit 2 incorporates the extended size of the source star, and Fit 3
  takes into account limb--darkening of the source. }
\tablenotetext{b} { JD - 2448623.50. }
\tablenotetext{c} { Einstein diameter crossing time. }
\tablenotetext{d} { Lens angular velocity, relative to the source, is in $\kmskpc$
    and assumes a $61 \pm 12 \rsun$ source at 9 kpc. }
\tablenotetext{} { Comparison of event parameters between microlensing fits for
  MACHO Alert 95--30.  Statistics are derived from simultaneous fits on
  all passbands.  Reported uncertainties in the final significant digit(s)
  are the maximum extent of the surface in parameter space which has a
  $\chi ^2$ greater than the best--fit value by 1. }
\end{deluxetable} 

\clearpage
\begin{deluxetable}{ccllllll}
\tablecaption{MACHO Alert 95-30 microlensing statistics \label{tab-stat2} }
\tablewidth{0pt}
\tablehead{
        \colhead {Passband} &
        \colhead {$\#$ Observations} &
        \colhead {$\bar{dm}$ \tablenotemark{a} } &
        \colhead {Fit 1 $\chi ^2$} &  
        \colhead {Fit 2 $\chi ^2$} &  
        \colhead {Fit 3 $\chi ^2$}    
}
\startdata
MACHO R  & 205             
         & 0.018
         & 980.33
         & 847.49
         & 839.15 \nl
MACHO V  & 292               
         & 0.023
         & 1069.65
         & 964.16
         & 963.86 \nl
CTIO R   & 103
         & 0.019
         & 307.58
         & 124.97
         & 126.51 \nl
MJUO R   & 41
         & 0.019
         & 184.56
         & 46.68
         & 42.62 \nl
UTSO V   & 55                
         & 0.020
         & 101.34
         & 48.64
         & 50.28 \nl
UTSO R   & 96                
         & 0.013
         & 488.20
         & 82.91
         & 90.09 \nl
WISE R   & 12                
         & 0.013
         & 80.71
         & 13.76
         & 7.22 \nl
\tableline
TOTAL    & 804
         &
         & 3212.40
         & 2128.64
         & 2119.76 \nl
\enddata
\tablenotetext{a} { Average error, in magnitudes, for each passband. }
\tablenotetext{} {Individual microlensing statistics for MACHO and
  GMAN observations of Alert 95--30.  The number of constraints per passband 
  for fits 1, 2, and 3 are four, five, and five, respectively. }
\end{deluxetable}

\clearpage
\begin{deluxetable}{llllllll}
\tablewidth{0pt}
\scriptsize
\tablecaption{Photometry of the source star in MACHO 95-30 \label{tab-phot1}}
\tablehead{
\multicolumn{1}{c}{Observed}&
\multicolumn{1}{c}{Extinction}&
\multicolumn{1}{c}{Dereddened}&
\multicolumn{1}{c}{Abs Mag, 8 kpc}&
\multicolumn{1}{c}{Abs Mag, 9 kpc}
}
\startdata
$V=16.21$ &$A_V=1.35$    &$V_0=14.86$ &$M_V=+0.34$&$M_V=+0.59$  \nl
$K=9.98$  &$A_K=0.15$    &$K_0=9.83$  &$M_K=-4.69$&$M_K=-4.45$ \nl
$V-R=1.39$&$E(V-R)=0.34$&$V-R_0=1.05$&          &       \nl
$J-K=1.12$&$E(J-K)=0.23$&$J-K_0=0.89$&          &       \nl
$H-K=0.26$&$E(H-K)=0.08$&$H-K_0=0.18$&          &       \nl
$V-K=6.23$&$E(V-K)=1.21$&$V-K_0=5.03$&          &       \nl
          &             &            &          &       \nl
Bolometric&             &$BC_K=-2.7\pm 0.1$&$\Mbol = -2.0$&$\Mbol = -2.25$  \nl
\enddata
\end{deluxetable}

\clearpage
\begin{deluxetable}{ccc} 
\tablecaption{Limb--darkening coefficients for the source star \label{tab-limbd} }
\tablewidth{0pt}
\tablehead{
        \colhead{Passband} &
        \colhead{a} &
        \colhead{b}
}
\startdata
MACHO V    & 1.140 
           & -0.284 \nl
MACHO R    & 0.825
           & -0.051 \nl
Standard V & 1.072 
           & -0.228 \nl
Standard R & 0.910 
           & -0.108 \nl
\enddata
\tablenotetext{}{Limb darkening coefficients for Equation~\ref{eq-limbd} used
  to approximate the brightness profile of the source star in MACHO Alert 95--30.}
\end{deluxetable}

\clearpage
\begin{deluxetable}{lllllll}
\small
\tablecaption{Summary of spectral observations appearing in Figure~\ref{fig-specdate} \label{tab-spec}}
\tablehead{
  \colhead{JD} &
  \colhead{Telescope \& Instrument} &
  \colhead{Dispersion} &
  \colhead{Resolution} &
  \colhead{Coverage} &
  \colhead{S / N}
}
\startdata
2449944.234 & CTIO 4m/RC Sp.    & 1.0 \AA\ ${\rm pix^{-1}}$  & 4.0\AA\ & 6230--9340\AA\ & $\sim 100$ \nl
2449945.211 & CTIO 4m/RC Sp.    & 1.0 \AA\ ${\rm pix^{-1}}$  & 4.0\AA\ & 6230--9340\AA\ & $\sim 100$ \nl
2449947.221 & Keck 10m/HIRES    & 0.04 \AA\ ${\rm pix^{-1}}$ & 0.2\AA\ & 4309--6739\AA\ & 92-256 \nl
2449947.524 & MSO 74"/Cass. Sp. & 0.9 \AA\ ${\rm pix^{-1}}$  & 4.6\AA\ & 6240--6770\AA\ & 106 \nl
2449948.223 & Keck 10m/HIRES    & 0.04 \AA\ ${\rm pix^{-1}}$ & 0.2\AA\ & 4835--7282\AA\ & 166-296 \nl
2449948.504 & MSO 74"/Cass. Sp. & 0.9 \AA\ ${\rm pix^{-1}}$  & 4.6\AA\ & 6240--6770\AA\ & 117 \nl
2449949.509 & MSO 74"/Cass. Sp. & 0.9 \AA\ ${\rm pix^{-1}}$  & 4.6\AA\ & 6240--6770\AA\ & 113 \nl
2449950.322 & Keck 10m/HIRES    & 0.04 \AA\ ${\rm pix^{-1}}$ & 0.2\AA\ & 3750--6065\AA\ & 20-141 \nl
2449950.523 & MSO 74"/Cass. Sp. & 0.9 \AA\ ${\rm pix^{-1}}$  & 4.6\AA\ & 6240--6770\AA\ &  91 \nl
2449951.498 & MSO 74"/Cass. Sp. & 0.9 \AA\ ${\rm pix^{-1}}$  & 4.6\AA\ & 6240--6770\AA\ &  96 \nl
2449952.508 & MSO 74"/Cass. Sp. & 0.9 \AA\ ${\rm pix^{-1}}$  & 4.6\AA\ & 6240--6770\AA\ &  72 \nl
2449953.494 & MSO 74"/Cass. Sp. & 0.9 \AA\ ${\rm pix^{-1}}$  & 4.6\AA\ & 6240--6770\AA\ &  78 \nl
2449954.513 & MSO 74"/Cass. Sp. & 0.9 \AA\ ${\rm pix^{-1}}$  & 4.6\AA\ & 6240--6770\AA\ &  66 \nl
2449987.000 & CTIO 4m/RC Sp.    & 2.0 \AA\ ${\rm pix^{-1}}$  & 8.0\AA\ & 3890--9830\AA\ & $\sim 100$\nl
\enddata
\end{deluxetable}


\clearpage
\begin{thebibliography}{10}

\bibitem[\protect\citeauthoryear{Abe \etal}
  {Abe \etal} {1997}]{moa}
  Abe,~F., \etal, 1997, in IAP Colloquium on ``Variable Stars and the Astrophysical Returns
  of Microlensing Surveys'', ed. R. Ferlet, in press

\bibitem[\protect\citeauthoryear{Alard \etal}{Alard \etal}{1995}]{duo}
  Alard,~C. \etal, 1995, ESO Messenger, 80, 31  
  
\bibitem[\protect\citeauthoryear{Alard, Mao, \& Guibert}
  {Alard, Mao, \& Guibert} {1995}]{duo2}
  Alard,~C., Mao,~S. \& Guibert,~J., 1995, \aap, 300, L17  
  
\bibitem[\protect\citeauthoryear{Alard \etal}{Alard \etal}{1996}]{alard-sgr}
  Alard,~C. \etal, 1996, \apj, 458, L17

\bibitem[\protect\citeauthoryear{Albrow \etal} 
  {Albrow \etal} {1997}] {planet}
  Albrow,~M. \etal, 1997, in IAP Colloquium on ``Variable Stars and the Astrophysical Returns
  of Microlensing Surveys'', ed. R. Ferlet, in press

\bibitem[\protect\citeauthoryear{Alcock \etal}
  {Alcock \etal} {1995a}]{macho-bulge1}
  Alcock,~C. \etal, 1995a, \apj, 445, 133
  
\bibitem[\protect\citeauthoryear{Alcock \etal}
  {Alcock \etal}{1995b}]{macho-parallax}
  Alcock,~C. \etal, 1995b, \apjl, 454, L125
  
\bibitem[\protect\citeauthoryear{Alcock \etal}
  {Alcock \etal} {1996a}]{macho-alert1}
  Alcock,~C. \etal, 1996a, \apjl, 463, L67
  
\bibitem[\protect\citeauthoryear{Alcock \etal}
  {Alcock \etal} {1996b}]{macho-bulge2}
  Alcock,~C. \etal, 1996b, \apj, submitted (astro-ph/9512146) 
  
\bibitem[\protect\citeauthoryear{Alcock \etal}
  {Alcock \etal} {1996c}]{macho-lmc2}
  Alcock,~C. \etal, 1996c, \apj, submitted (astro-ph/9606165) 
  
\bibitem[\protect\citeauthoryear{Alcock \etal}
  {Alcock \etal} {1996d}] {macho-causitc}
  Alcock,~C. \etal, 1996d, \iaucirc, 6361

\bibitem[\protect\citeauthoryear{Alcock \etal}
  {Alcock \etal} {1997}] {macho-sgr}
  Alcock,~C. \etal, 1997, \apj, 474, 217

\bibitem[\protect\citeauthoryear{Alloin \& Bica}
  {Alloin \& Bica}{1989}] {alloin}
  Alloin,~D.~M., \& Bica,~E., 1989, \aap, 217, 57

\bibitem[\protect\citeauthoryear{Aubourg \etal}
  {Aubourg \etal} {1995}] {eros-ccd}
  Aubourg,~E. \etal, 1995, \aap, 301, 1 
  
\bibitem[\protect\citeauthoryear{Barbuy \etal}
  {Barbuy \etal} {1992}] {barbuy}
  Barbuy,~B., Castro,~S., Ortolani,~S., \& Bica,~E., 1992, \aap, 259, 607
  
\bibitem[\protect\citeauthoryear{Benetti \etal}
  {Benetti \etal} {1995}] {benetti}
  Benetti,~S., Pasquini,~L., \& West,~R.~M., 1995, \aap, 294, L37

\bibitem[\protect\citeauthoryear{Bennett \& Rhie}
  {Bennett \& Rhie}{1996}] {ben-rhie}
  Bennett,~D.~P., \& Rhie,~S.~H., 1996, \apj, 472, 660

\bibitem[\protect\citeauthoryear{Bessell \& Wood}
  {Bessell \& Wood}{1984}] {bessell}
  Bessell,~M.~S., \& Wood,~P.~R. 1984, \pasp, 96, 247

\bibitem[\protect\citeauthoryear{Bessell \etal}
  {Bessell \etal} {1997}] {bessell2}
  Bessell,~M.~S., Castelli,~F., \& Plez,~B.  1997, preprint

\bibitem[\protect\citeauthoryear{Binney \etal}
  {Binney \etal} {1996}] {binney}
  Binney,~J.~J., Gehrard,~O.~E., \& Spergel,~D., 1996, \mnras, in press (astro-ph/9609066)

\bibitem[\protect\citeauthoryear{Bono \etal}
  {Bono \etal} {1997}] {bono}
  Bono,~G., Caputo,~F., Castellani,~V., \& Marconi,~M., 1997, \aap, in press

\bibitem[\protect\citeauthoryear{Carney \etal}
  {Carney \etal}{1995}] {carney}
  Carney,~B.~W., Fulbright,~J.~P., Terndrup,~D.~M., Suntzeff,~N.~B.,
  \& Walker,~A.~R., 1995, \aj, 110, 1674

\bibitem[\protect\citeauthoryear{Claret, D{\'i}az-Cordov{\'e}s, \& Gim{\'e}nez}
  {Claret \etal} {1995}] {claret-r}
  Claret,~A., D{\'i}az-Cordov{\'e}s,~J., \& Gim{\'e}nez,~A., 1995, \aaps, 114, 247
  
\bibitem[\protect\citeauthoryear{Claret}{Claret} {1996}] {claret-macho}
  Claret,~A., 1996, Private communication
  
\bibitem[\protect\citeauthoryear{Crotts}
  {Crotts}{1996}] {vatt}
  Crotts,~A.~P.~S., 1996, astro-ph/9610067

\bibitem[\protect\citeauthoryear{Diaz \etal}
  {Diaz \etal} {1989}] {diaz}
  Diaz,~A., Terlevich,~E., \& Terlevich,~R., 1989, \mnras, 239, 325
  
\bibitem[\protect\citeauthoryear{D{\'i}az-Cordov{\'e}s, Claret, \& Gim{\'e}nez}
  {D{\'i}az-Cordov{\'e}s \etal} {1995}] {claret-v}
  D{\'i}az-Cordov{\'e}s,~J., Claret,~A., \& Gim{\'e}nez,~A., 1995, \aaps, 110, 329
  
\bibitem[\protect\citeauthoryear{Di Stefano \& Esin}
  {Di Stefano \& Esin} {1995}] {distefano95}
  Di Stefano,~R. \& Esin,~A.~A., 1995, \apjl, 448, L1
  
\bibitem[\protect\citeauthoryear{Dyck \etal}
  {Dyck \etal}{1996}] {dyck}
  Dyck,~H.~M., Benson,~J.~A., van~Belle,~G.~T., \& Ridgway,~S.~T.,
  1996, \aj, 111, 1705

\bibitem[\protect\citeauthoryear{Erdelyi-Mendez \& Barbuy}
  {Erdelyi-Mendez \& Barbuy} {1991}] {erdelyi}
  Erdelyi-Mendez,~M., \& Barbuy,~B., 1991, \aap, 241, 176

\bibitem[\protect\citeauthoryear{Feast}
  {Feast} {1996}] {feast}
  Feast,~M.~W., 1996, \mnras, 278, 11

\bibitem[\protect\citeauthoryear{Frogel \& Whitford}
  {Frogel \& Whitford} {1987}] {frogel-whit}
  Frogel,~J. A., \& Whitford,~A.~E., 1987, \apj, 320, 199

\bibitem[\protect\citeauthoryear{Frogel \etal}
  {Frogel \etal} {1981}] {frogel}
  Frogel,~J.~A., Persson,~S.~E., \& Cohen,~J.~G., 1981, \apj, 246, 842

\bibitem[\protect\citeauthoryear{Gondolo \etal}
  {Gondolo \etal} {1996}] {agape}
  Gondolo,~P., \etal, 1996, astro--ph/9610010

\bibitem[\protect\citeauthoryear{Gould}{Gould} {1994}] {gould-pm}
  Gould,~A., 1994, \apj, 421, L71  
  
\bibitem[\protect\citeauthoryear{Gould}{Gould} {1996}] {gould-rev}
  Gould,~A., 1996, \pasp, 108, 465 
  
\bibitem[\protect\citeauthoryear{Gould}{Gould} {1997}] {gould-rot}
  Gould,~A., 1997, \apj, Submitted
  
\bibitem[\protect\citeauthoryear{Gould \& Loeb}
     {Gould \& Loeb} {1992}] {gould-loeb}
 Gould,~A. \& Loeb,~A., 1992, \apj, 396, 104  

%
\bibitem[\protect\citeauthoryear{Gould \& Welch}
  {Gould \& Welch} {1996}] {gould-welch}
  Gould,~A. \& Welch,~D.~L., 1996, \apj, 464, 212  

\bibitem[\protect\citeauthoryear{Griest \etal}
  {Griest \etal}  {1991}] {macho-griest91}
  Griest,~K. \etal, 1991, \apjl, 372, L79   
  
\bibitem[\protect\citeauthoryear{Guarnieri \etal}
  {Guarnieri \etal} {1996}] {guarnieri}
  Guarnieri,~D., \etal, 1996, in ``Science with the HST II'', eds. P. Benvenutti, D. Macchetto,
  \& E. Schreier (STScI: Baltimore), p. 338

\bibitem[\protect\citeauthoryear{Han \& Gould}
  {Han \& Gould} {1995}] {han-bar}
  Han,~C. \& Gould,~A., 1995, \apj, 447, 53

\bibitem[\protect\citeauthoryear{Han \& Gould}
  {Han \& Gould} {1996}] {han-spec}
  Han,~C. \& Gould,~A., 1996, \apj, 467, 540

\bibitem[\protect\citeauthoryear{Holtzman \etal}
  {Holtzman \etal} {1993}] {holtzman}
  Holtzman,~J., \etal, 1993, \aj, 106, 1826
  
\bibitem[\protect\citeauthoryear{Humphreys \& Graham}
  {Humphreys \& Graham} {1986}] {humphreys}
  Humphreys,~R.~M., \& Graham,~J.~A., 1986, \aj, 91, 522
  
\bibitem[\protect\citeauthoryear{Ibata \etal}
  {Ibata \etal} {1995}] {ibata}
  Ibata,~R., Gilmore,~G., \& Irwin,~M.~J., 1995, Nature, 370, 194
  
\bibitem[\protect\citeauthoryear{James}{James} {1994}] {minuit}
  James,~F., 1994, CERN Program Library Long Writeup D506.
  
\bibitem[\protect\citeauthoryear{Jones \etal}
  {Jones \etal} {1984}] {jones}
  Jones,~J.~E., Alloin,~D.~M., \& Jones,~B.~T., 1984, \apj, 283, 457
  
\bibitem[\protect\citeauthoryear{Jorgensen \etal}
  {Jorgensen \etal} {1992}] {jorgensen}
  Jorgensen,~U.~G., Carlsson,~M., \& Johnson,~H.~R., 1992, \aap, 254, 258
  
\bibitem[\protect\citeauthoryear{Kiraga \& Paczy{\'n}ski}
  {Kiraga \& Paczy{\'n}ski} {1994}] {kir-pac}
  Kiraga,~M., \& Paczy{\'n}ski,~B., 1994, \apjl, 430, L101

\bibitem[\protect\citeauthoryear{Kirkpatrick \etal} 
  {Kirkpatrick \etal} {1991}] {kirkpatrick}
  Kirkpatrick,~J.~D., Henry,~T.~J., \& McCarthy,~D.~W., 1991, \apjs, 77, 417
  
\bibitem[\protect\citeauthoryear{Lang}{Lang} {1992}] {lang}
  Lang, K. R. 1992, Astrophysical Data (New York: Springer-Verlag)
  
\bibitem[\protect\citeauthoryear{Loeb \& Sasselov}
  {Loeb \& Sasselov} {1995}]{loeb-sass}
  Loeb,~A., \& Sasselov,~D., 1995, \apjl, 449, L33

\bibitem[\protect\citeauthoryear{Mao \& Paczy{\'n}ski}
  {Mao \& Paczy{\'n}ski} {1991}] {mao-pac}
  Mao,~S. \& Paczy{\'n}ski,~B., 1991,  \apjl, 374, L37    

\bibitem[\protect\citeauthoryear{Maoz \& Gould}
  {Maoz \& Gould} {1994}] {maoz-gould}
  Maoz,~D., \& Gould,~A., 1994, \apjl, 425, L67

\bibitem[\protect\citeauthoryear{Minniti}
  {Minniti} {1996}] {minniti}
  Minniti,~D., 1996, \apj, 459, 579

\bibitem[\protect\citeauthoryear{Minniti \etal}
  {Minniti \etal} {1996}] {minniti-liebert}
  Minniti,~D., Liebert,~J.~W., Olszewski,~E.~W., \& White,~S.~D.~M., 1996, \aj, 112, 590

\bibitem[\protect\citeauthoryear{Nemiroff \& Wickramasinghe}
  {Nemiroff \& Wickramasinghe} {1994}] {nemiroff94}
  Nemiroff,~R.~J. \& Wickramasinghe,~W.~A.~D.~T., 1994, \apjl, 424, L21
  
\bibitem[\protect\citeauthoryear{Ortolani \etal}
  {Ortolani \etal} {1990}] {ortolani}
  Ortolani,~S., Barbuy,~B., \& Bica,~E., 1990, \aap, 236, 362
  
\bibitem[\protect\citeauthoryear{Paczy{\'n}ski} 
  {Paczy{\'n}ski}  {1991}] {pac91}
  Paczy{\'n}ski,~B, 1991, \apjl, 371, L63   
  
\bibitem[\protect\citeauthoryear{Paczy{\'n}ski} 
  {Paczy{\'n}ski}  {1996}] {pac-rev}
  Paczy{\'n}ski,~B, 1996, \araa, In press
  
\bibitem[\protect\citeauthoryear{Paczy{\'n}ski \etal} 
  {Paczy{\'n}ski \etal} {1995}] {ogle}
  Paczy{\'n}ski,~B, \etal, 1995, \baas, 187, \# 14.07

\bibitem[\protect\citeauthoryear{Peng}
  {Peng} {1997}] {peng}
  Peng,~E.~W., 1997, \apj, 475, 43

\bibitem[\protect\citeauthoryear{Pratt \etal} 
  {Pratt \etal} {1995}] {macho-pratt}
  Pratt,~M.R. \etal, 1995, 
  In {\em Astrophysical Applications of Gravitational Lensing},
  IAU Symp. 173,  eds. Kochanek,~C.S. \& Hewitt,~J.N, Kluwer. 
  (astro-ph/9508039)
  
\bibitem[\protect\citeauthoryear{Renault \etal}
  {Renault \etal} {1996}] {eros}
  Renault.~C, \etal, 1996, \aap, submitted (astro-ph/9612102)

\bibitem[\protect\citeauthoryear{Rieke \& Lebofski}
  {Rieke \& Lebofski} {1985}] {rieke}
  Rieke,~G., \& Lebofski,~M., 1985, \apj, 288, 618
  
\bibitem[\protect\citeauthoryear{Sasselov}
  {Sasselov} {1997}]{sasselov}
  Sasselov,~D., 1997, in IAP Colloquium on ``Variable Stars and the Astrophysical Returns
  of Microlensing Surveys'', ed. R. Ferlet, in press

\bibitem[\protect\citeauthoryear{Stetson}{Stetson} {1994a}] {daophot}
  Stetson,~P.~B., 1994a, Daophot II User's manual.
  
\bibitem[\protect\citeauthoryear{Stetson}{Stetson} {1994b}] {allframe}
  Stetson,~P.~B., 1994b, \pasp, 106, 250
  
\bibitem[\protect\citeauthoryear{Turnshek \etal}
  {Turnshek \etal} {1985}] {turnshek}
  Turnshek,~D.~E., Turnshek,~D.~A., Craine,~E.~R., \& Boesjaar,~P.~C., 1985, 
  An Atlas of Digital Spectra of Cool Stars (Tucson: Western Research Co.)
  
\bibitem[\protect\citeauthoryear{Udalski \etal} 
  {Udalski \etal} {1994a}]{ogle-tau}
  Udalski,~A., Szyma{\'n}ski,~M., Kalu{\.z}ny,~J., Kubiak,~M., 
  Krzemi{\'n}ski,~W., Mateo,~M.,  Preston,~G.~W., \& Paczy{\'n}ski,~B., 1994a, 
  \newblock {\rm Acta Astronomica}, 44, 165 
  
\bibitem[\protect\citeauthoryear{Udalski \etal}
  {Udalski \etal} {1994b}] {ogle7}
  Udalski,~A., Szyma{\'n}ski,~M., Mao,~S., Di Stefano,~R., 
  Kalu{\.z}ny,~J., Kubiak,~M., Mateo,~M. \& Krzemi{\'n}ski,~W., 1994b, 
  \apjl, 436, L103     
  
\bibitem[\protect\citeauthoryear{Witt \& Mao}
  {Witt \& Mao} {1994}] {witt-mao}
  Witt,~H.~J., \& Mao,~S., 1994, \apj, 430, 505   

\bibitem[\protect\citeauthoryear{Zhao \etal}
  {Zhao \etal} {1996}] {bar}
  Zhao,~H., Rich,~R.~M., \& Spergel,~D.~N., 1996, \mnras, 282, 175

\end{thebibliography}
\end{document}